\documentclass[twocolumn]{article}
\usepackage{graphicx} 
\usepackage{amsmath}
\usepackage{amssymb}
\usepackage{caption}
\usepackage{subcaption}
\usepackage{siunitx}
\usepackage[hyphens]{url}
\usepackage{hyperref}
\usepackage{float}
\hypersetup{breaklinks=true}
\usepackage{tabularx}
\usepackage{listings}
\begin{document}

\title{Vertex Shader Domain Warping with Automatic Differentiation}

\author
       {Dave Pagurek van Mossel\\Butter Creatives
       }



\twocolumn[{
\maketitle
\begin{center}
    \captionsetup{type=figure}
    \includegraphics[trim={5cm 1cm 5cm 10cm},clip,width=1.2in]{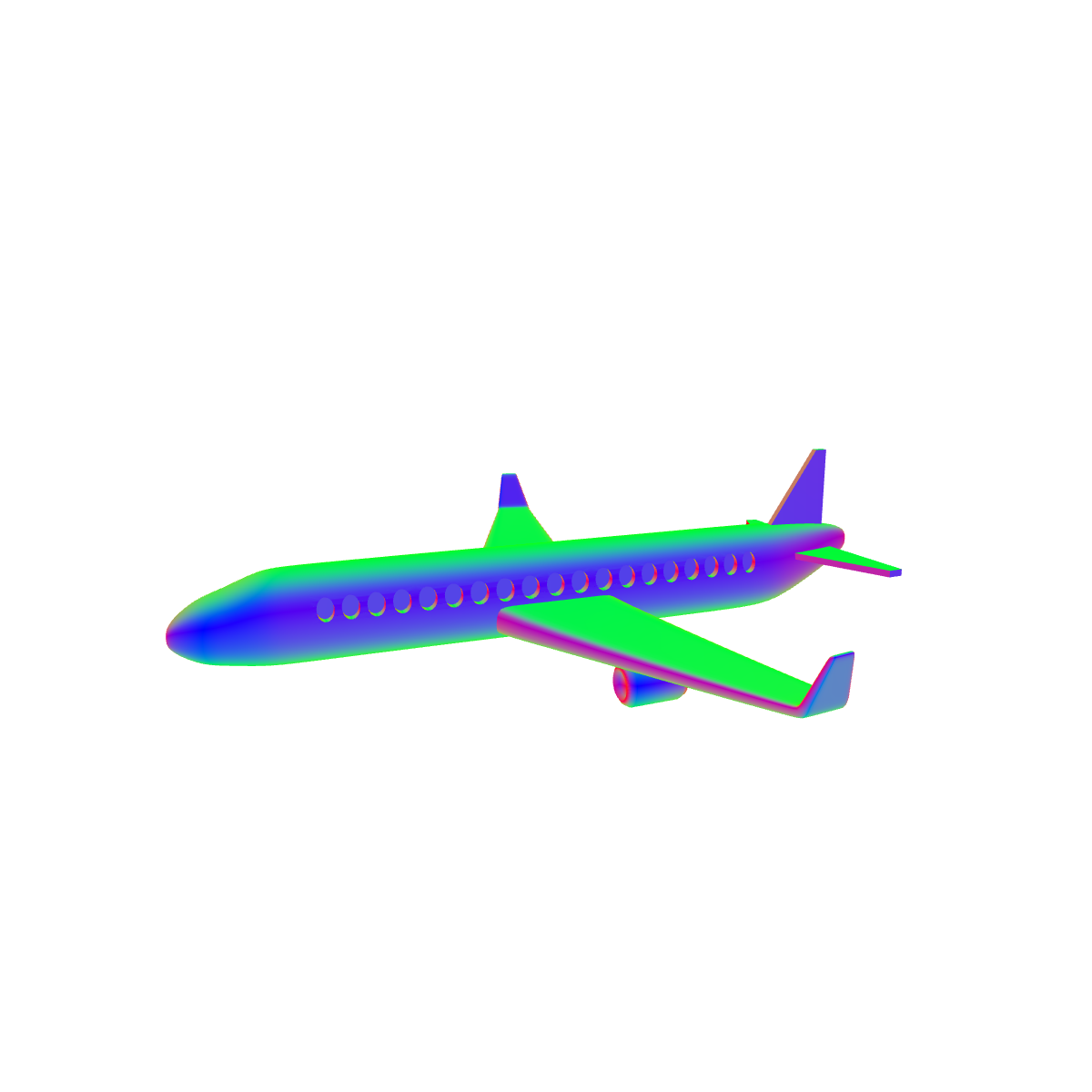}
    \includegraphics[trim={5cm 1cm 5cm 10cm},clip,width=1.2in]{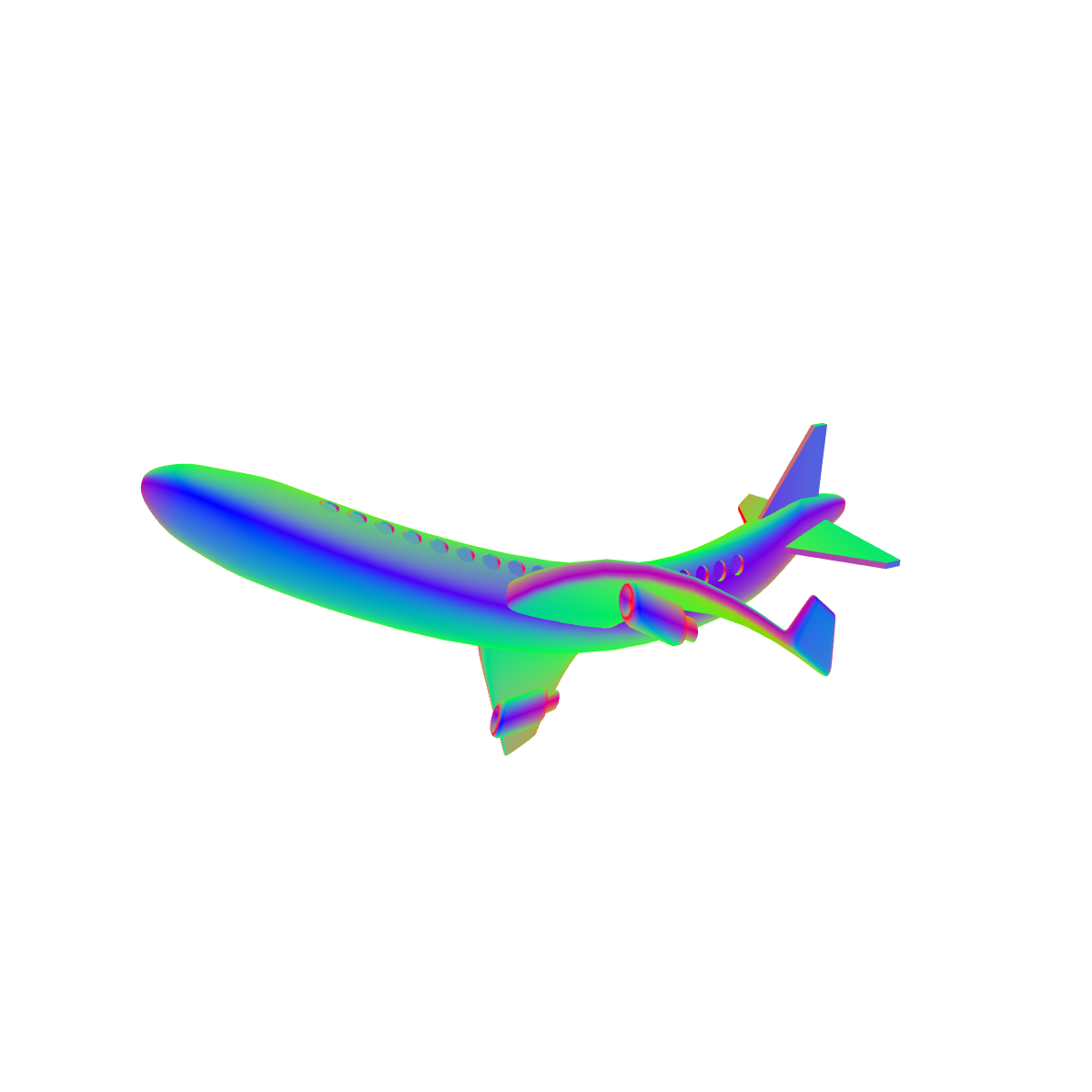}
    \includegraphics[trim={5cm 1cm 5cm 10cm},clip,width=1.2in]{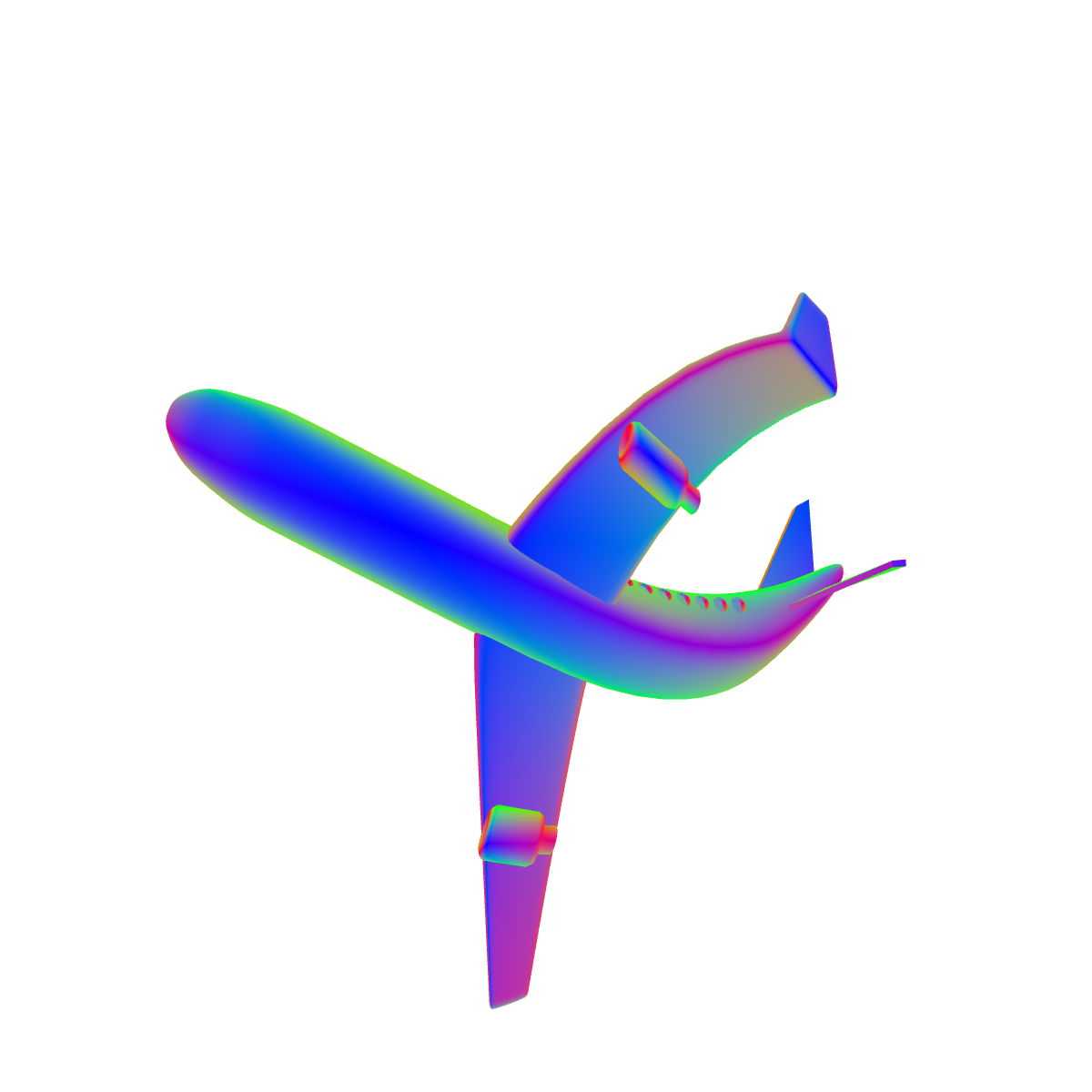}
    \includegraphics[trim={5cm 1cm 5cm 10cm},clip,width=1.2in]{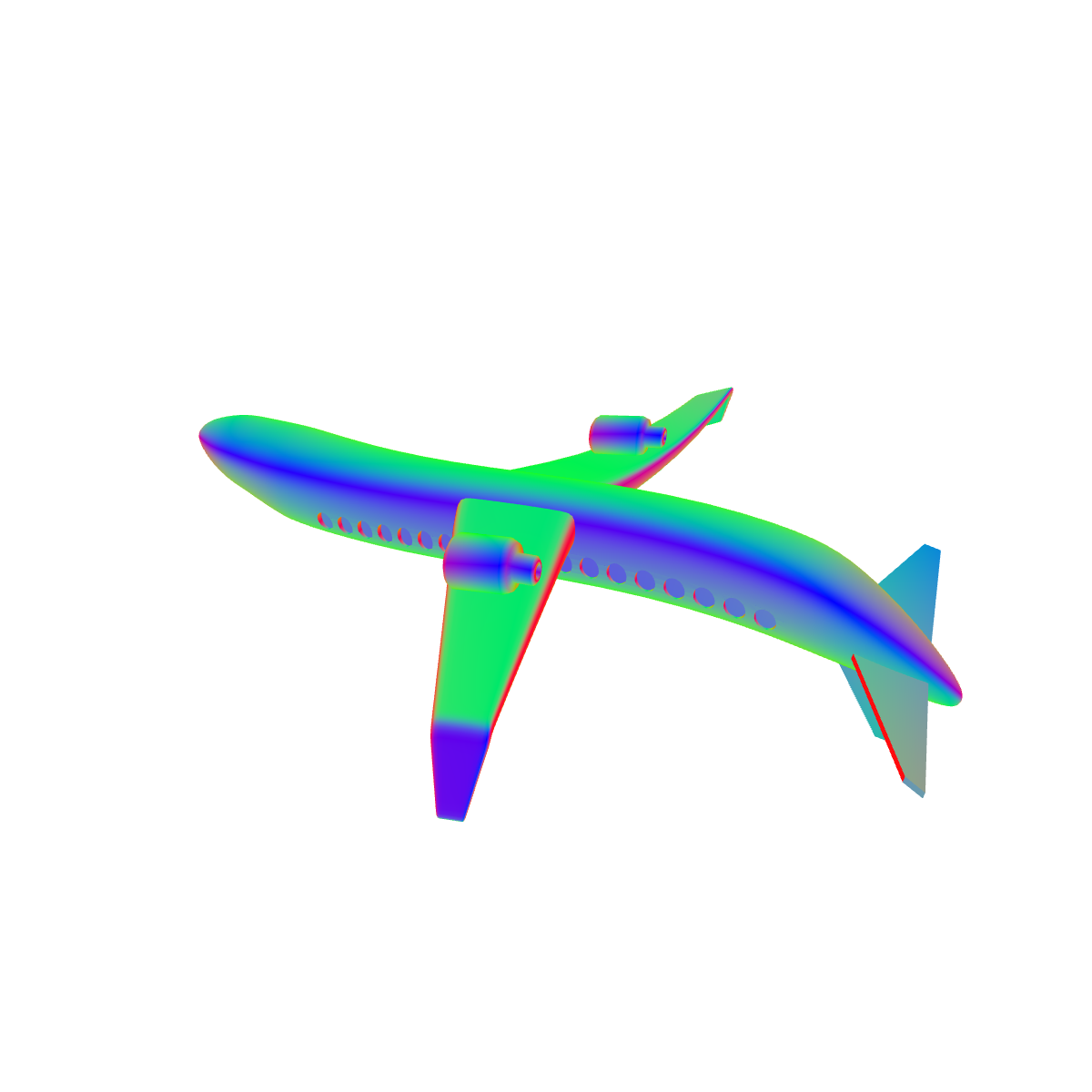}
    \captionof{figure}{An airplane model undergoing an animated twist warp. Its vertex positions and normals are updated in the vertex shader, with the fragment shader visualizing the absolute normal direction as a color. After applying the warp, the output normals are accurate.}
  \label{fig:teaser}
\end{center}
}]

\begin{abstract}
\small
Domain warping is a technique commonly used in creative coding to distort graphics and add visual interest to a work. The approach has the potential to be used in 3D art as mesh vertices can be efficiently warped using a vertex shader in a WebGL pipeline. However, 3D models packaged for the web typically come with baked-in normal vectors, and these need to be updated when vertex positions change for lighting calculations to work. This is typically done via finite differences, which requires parameter tuning to achieve optimal visual fidelity. We present a method for 3D domain warping that works with automatic differentiation, allowing exact normals to be used without any tuning while still benefiting from hardware acceleration.
\end{abstract}


\section{Introduction}
\label{sec:introduction}
Creative coding is an art form that uses the computer and source code as an expressive medium. The process of writing creative code is often characterized by exploration and iteration (projects made in Processing~\cite{Processing} are called ``sketches'' for this reason~\cite{FryThesis}) and thus many commonly employed techniques are ones with easily tweakable parameters that can generate a wide variety of visual results.


One such technique is \textit{domain warping}, used often in 2D and some 3D contexts~\cite{DomainWarping}. Given a function $f:~\mathbb{R}^n~\mapsto~\mathbb{R}^n$ that defines an offset for any point in space, using the warp function $w(\vec{x}) = \vec{x} + f(\vec{x})$, one can warp any function whose domain is $\mathbb{R}^n$ by giving it $w(\vec{x})$ as input instead of $\vec{x}$. It provides a framework that invites exploration, as any offset function can be used, producing varied, interesting results from combinations of simple mathematical building blocks. While some carefully tuned offset functions may have physical motivations, such as the wave equations of Dynamic Kelvinlets~\cite{DynamicKelvinlets}, this is not a requirement. Arbitrary functions are useful tools in the creation of surreal or abstract art, a common style in creative coding~\cite{MattdslWarp,KovachWarp,GrewebWarp}.

\begin{figure*}
  \centering
    \begin{subfigure}[t]{0.21\textwidth}
        \includegraphics[trim={0cm 0cm 0cm 0cm},clip,width=\textwidth]{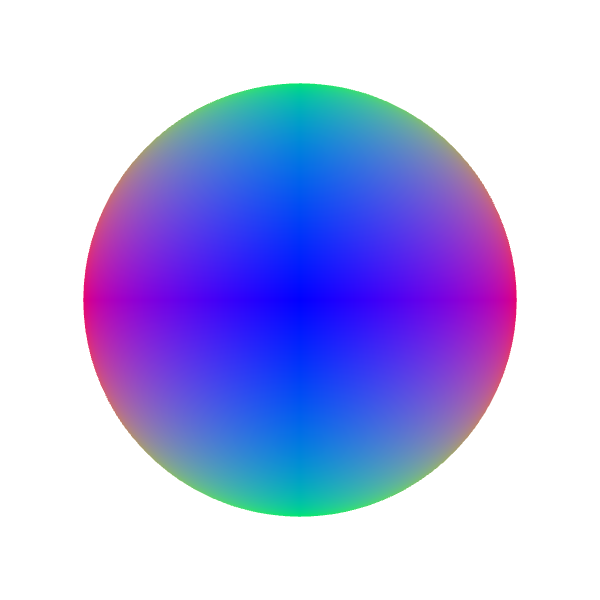}
        \caption{Input}
    \end{subfigure}
    \hspace{1em}
    \begin{subfigure}[t]{0.21\textwidth}
        \includegraphics[trim={0cm 0cm 0cm 0cm},clip,width=\textwidth]{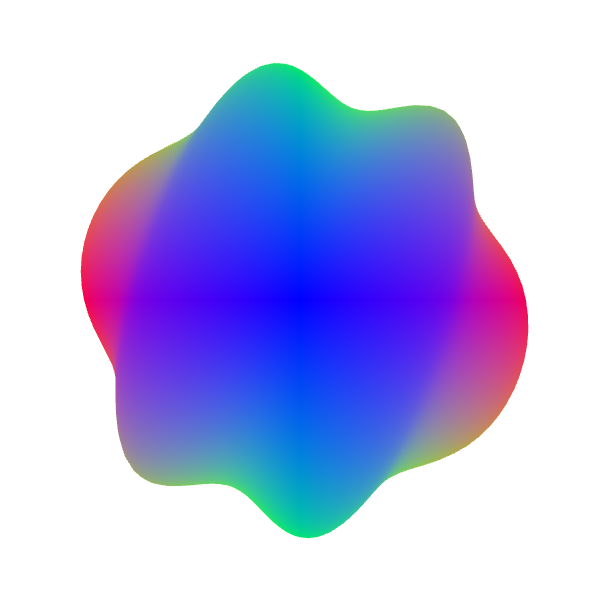}
        \caption{\label{fig:normals-none}Warp without adjusting normals}
    \end{subfigure}
    \hspace{1em}
    \begin{subfigure}[t]{0.21\textwidth}
        \includegraphics[trim={0cm 0cm 0cm 0cm},clip,width=\textwidth]{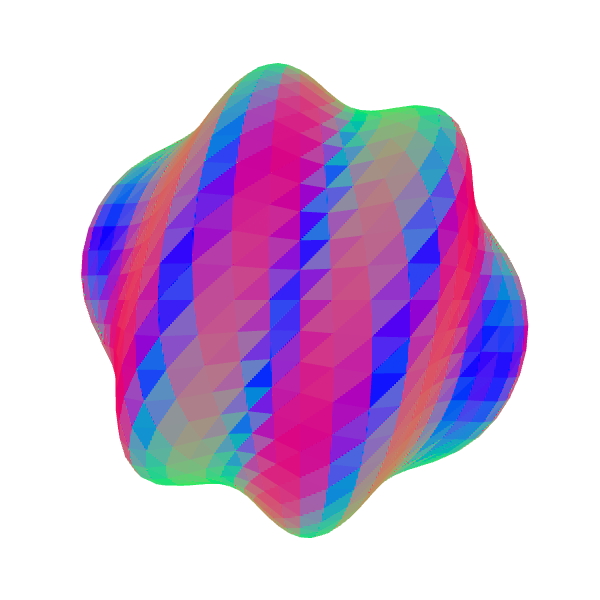}
        \caption{\label{fig:normals-ss}Warp with normals from screen-space derivatives}
    \end{subfigure}
    \hspace{1em}
    \begin{subfigure}[t]{0.21\textwidth}
        \includegraphics[trim={0cm 0cm 0cm 0cm},clip,width=\textwidth]{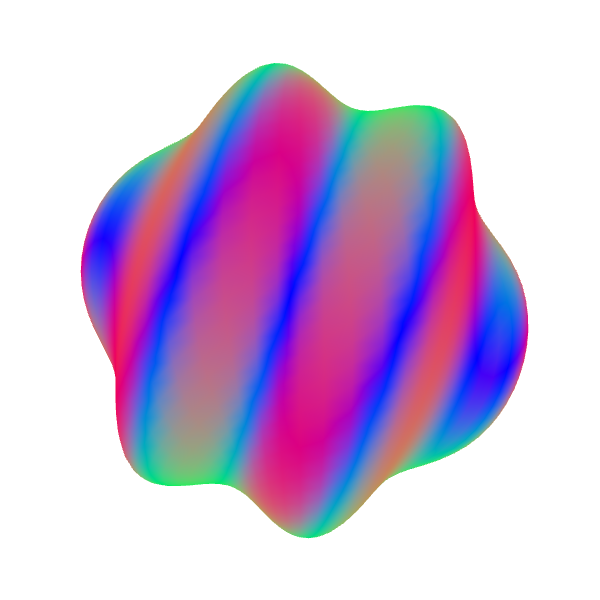}
        \caption{\label{fig:spherewarp}Warp affecting vertices and normals}
    \end{subfigure}
  
   \caption{\label{fig:normals}
     When applying a sine wave warp to the vertices of a sphere~(a), the baked normals will be incorrect~(b) unless they are updated to account for the warp. Relying on screen-space derivatives of position yields faceted normals~(c), while computing normals in the vertex shader allows for smooth normals~(d).}
\end{figure*}

This technique has the potential to fit nicely into a 3D workflow on the web: to warp the domain of a 3D mesh, one can implement an offset function in a vertex shader, which is the step in the pipeline responsible for transforming vertex positions into screen space before rasterization. Running on the GPU, this is likely the most efficient method of domain warping a triangle mesh. Unfortunately, meshes packaged for a WebGL pipeline typically bake in a normal vector for each vertex. If one applies domain warping to vertex positions without updating the baked normals, they will no longer be perpendicular to the surface of the mesh (visualized in Figure~\ref{fig:normals-none}) and will cause later lighting calculations in the fragment shader to be inaccurate. Using finite differences to approximate updated normals requires parameter tuning for each use to avoid visual artifacts, making it difficult to use in a general system accepting arbitrary meshes and offset functions. We present an algorithm that leverages automatic differentiation to generate an updated normal in any circumstance without parameter tuning, enabling 3D domain warping to be used more easily in web-based creative code environments.

\section{Background}
\label{sec:background}

\paragraph{Domain warping.} Domain warping in 3D is a common technique used when raymarching signed distance functions (SDFs)~\cite{DomainWarping}, found often in the demoscene and on platforms like Shadertoy~\cite{Shadertoy}. Unlike triangle meshes, SDFs do not have baked normals that need updating: models are represented by a function $f:~\mathbb{R}^3~\mapsto~\mathbb{R}$ describing, in theory, the distance to the surface of the shape at a given point in space, and in practice, a conservative estimate of said distance. Shapes are then rendered by marching (or ``sphere tracing'') light rays through the scene, detecting intersections using the scene SDF~\cite{Raymarching}. While domain warping an inexact SDF can be as simple as function composition, one cannot make use of existing triangle-based 3D models without a nontrivial conversion that comes with performance tradeoffs~\cite{mesh2sdf}.

\paragraph{Perturbing normals.} In a mesh-based WebGL pipeline, perturbations that update normals are used at small scales to add extra texture to the surface of a model. Traditionally, this technique is called \textit{bump mapping} and involves parameterizing the surface of a model into two axes, $u$ and $v$, and providing a bump height $h:~\mathbb{R}^2~\mapsto\mathbb{R}$ defined for each parameter value~\cite{bumpmapping}. The bump map suggests a new surface, where each point is moved along its normal by the height defined by the bump function, yielding $\vec{p'} = \vec{p} + h([u_p~~v_p]^T)\hat{n}$. One can then compute a perturbed normal $\hat{n}' = \frac{\partial \vec{p'}}{\partial u} \times \frac{\partial \vec{p'}}{\partial v}$. This formulation does not depend on a specific parameterization of the surface, so bump maps can be applied in a fragment shader by using a local screen-space parameterization~\cite{Mikkelsen2010BumpMU}, relying on derivatives with respect to neighboring pixels (\lstinline{dFdx(position)} and \lstinline{dFdy(position)} in GLSL.)

Bump mapping intentionally only affects surface lighting and not the geometry, and given this small scale, only handles 1D height maps. If one wants to apply similar techniques at larger scales using 3D offsets, updated formulae are required. At larger scales, the vertices must be updated in the vertex shader, where one does not have screen-space derivatives available. If one were to recalculate normals using screen-space derivatives in the fragment shader based on the updated vertex positions, the linear interpolation between vertices becomes apparent, making normals appear faceted instead of smooth, shown in Figure~\ref{fig:normals-ss}.

\paragraph{Finite differences.} SDF raymarching code typically calculates normals via finite differences, approximating the normal ($\nabla f(\vec{p})$ when $\vec{p}$ is on the surface) with, using the X axis as an example, $\frac{1}{h}\left(f(\vec{p}+[h~0~0]^T)-f(\vec{p})\right)$ for a small value $h$. While this method of calculating normals can also work in a mesh-based WebGL pipeline, it requires simple but tedious tuning of $h$ to work well. To paint a picture of what picking $h$ involves, its value needs to be small enough to not lose detail, but not so small that numerical precision issues become visible. Since SDF raymarching is done per pixel, to be as large as possible without losing visual detail, a recommended heuristic for picking $h$ is to try to make its footprint in screen space approximately one pixel: small enough that more accurate normals would not be visible, and large enough to prevent aliasing.~\cite{NormalsSDF}

In the demoscene community, which produced many contemporary SDF techniques, file size constraints are common (including the entire genre of 64K intros), so fine-tuning of parameters is a common and acceptable tradeoff for smaller code. Our method, aimed instead at an audience interested in easy experimentation and expression, sidesteps this with a setup that leverages exact derivatives. Comparatively, normals achieved through automatic differentiation will be more accurate while having equivalent computational cost~\cite{AD}, motivating our approach and its use of automatic differentiation instead of approximation.

\section{Method}
\label{sec:formulation}

The artist provides an offset function $f:~\mathbb{R}^3~\mapsto~\mathbb{R}^3$ defining, for an input point in 3D space, a 3D offset to add to its location. We assume $f$ is continuous and differentiable. Each vertex position $\vec{p}$ on the mesh $M$ will be mapped to an updated position $\vec{p'} = w(\vec{p}) = \vec{p} + f(\vec{p})$.

Each $\vec{p}$ from the input mesh comes with a corresponding baked normal $\hat{n}$. We are solving for the vector $\hat{n}'$, representing the surface normal of the deformed mesh $M'$ at point $\vec{p'}$. This must be done in a way that exact derivatives that can be feasibly calculated in a shader, with only knowledge of per-vertex properties and not the whole mesh.

Our method achieves these goals in the following steps:
\begin{enumerate}
    \item We express derivatives of the warp function given a local tangent and bitangent, which we then use to generate a warped normal. Importantly, the structure of the derivative formula depends only on the warp function and accepts \textit{any} local parameterization, enabling one warp shader to work on all meshes.
    \item We provide a formula for a surface tangent and bitangent given only the surface normal, allowing inputs to Step~1 to be generated in a vertex shader with limited knowledge of the mesh.
    \item We describe a system to use automatic differentiation to statically generate a shader with the derivatives required for Step~1.
\end{enumerate}

\subsection{Updated Normals}
\label{sec:updatednormals}

Assume one has two unique vectors $\hat{u}$ and $\hat{v}$ tangent to the surface of the input mesh $M$ at $p$ such that $\hat{u} \times \hat{v} = \hat{n}$. One can approximate $\hat{n}'$ by finding the result of $w$ on points near $\vec{p}$, shifted slightly along a linear approximation of the surface at $\vec{p}$ by a small value $h$:
\begin{align*}
&\hat{n}' \approx\\
&\operatorname{normalize}\left(\frac{w(\vec{p} + h\hat{u}) - w(\vec{p})}{h} \times \frac{w(\vec{p} + h\hat{v}) - w(\vec{p})}{h}\right)
\end{align*}

Taking the limit as $h \rightarrow 0$, this is equivalent to taking directional derivatives of $w(\vec{p})$ in the directions $\hat{u}$ and $\hat{v}$:
\begin{align*}
    \hat{n}' &= \operatorname{normalize}\left(\frac{\partial w(\vec{p})}{\partial \hat{u}} \times \frac{\partial w(\vec{p})}{\partial\hat{v}}\right)\\
    &= \operatorname{normalize}\left(\frac{\partial (\vec{p} + f(\vec{p}))}{\partial \hat{u}} \times \frac{\partial (\vec{p} + f(\vec{p}))}{\partial\hat{v}}\right)
\end{align*}

Since, by definition, $\hat{u}$ and $\hat{v}$ are tangent to the surface at $\vec{p}$, the directional derivative of $\vec{p}$ in the direction of $\hat{u}$ or $\hat{v}$ will be $\hat{u}$ or $\hat{v}$ itself, respectively:
\begin{align*}
    \hat{n}' &= \operatorname{normalize}\left(\left(\hat{u} + \frac{\partial f(\vec{p})}{\partial \hat{u}}\right) \times \left(\hat{v} + \frac{\partial f(\vec{p})}{\partial\hat{v}}\right)\right)
\end{align*}

Since $f$ is defined in terms of Cartesian coordinates and not the coordinate space defined by $\hat{u}$ and $\hat{v}$, we reframe the above expression to be in terms of the Jacobian of the offset, $J=\left[\frac{\partial f(\vec{p'})}{\partial x}~\frac{\partial f(\vec{p'})}{\partial y}~\frac{\partial f(\vec{p'})}{\partial z}\right]$:
\begin{align}
    \hat{n}' &= \operatorname{normalize}\left(\left(\hat{u} + J\hat{u}\right) \times \left(\hat{v} + J\hat{v}\right)\right)\label{eqn:n}
\end{align}

To define $J$, we calculate all the partial first-order derivatives of $f$, which will be implemented using automatic differentiation, described later in Section~\ref{sec:ad}. Importantly, this is independent of the choice of tangent vectors: whichever ones we pick, the formula for $J$ does not change, so we do not need to run automatic differentiation at runtime. It is sufficient to run it once at compile time, and our runtime choice of tangent vectors only adds a matrix multiplication with $J$.

\subsection{Picking Tangent Vectors}

Without knowledge of the rest of the mesh vertices and their connectivity, we can make assumptions about the local surface around $\vec{p}$. The normal $\hat{n}$ implies that there is a surface plane $S$ passing through $\vec{p}$ and normal to $\hat{n}$. We can parameterize this plane into a tangent $\hat{u}$ and bitangent $\hat{v}$ by finding any vector $\hat{w}$ not equal to $\pm\hat{n}$, taking the cross product between it and $\hat{n}$ to get one vector parallel to $S$, and then taking the cross product between that and $\hat{n}$ to get another, different vector parallel to $S$:
\begin{align}
    \hat{w} &= \begin{cases}
        [0~~1~~0]^T,&\hat{n} = [\pm1~~0~~0]^T\nonumber\\
        [1~~0~~0]^T&\text{otherwise}\nonumber
    \end{cases}\nonumber\\
    \hat{v} &= \operatorname{normalize}(\hat{v} \times \hat{n})\label{eqn:v}\\
    \hat{u} &= \hat{v} \times \hat{n}\label{eqn:u}
\end{align}

Since $|\vec{a}\times\vec{b}|=|\vec{a}||\vec{b}|\sin\theta$, to get a \textit{unit} tangent vector from a cross product, one must normalize the result if the two inputs are not orthogonal. Since $\hat{w}$ and $\hat{v}$ are guaranteed by construction to not be parallel but may not be orthogonal, we must normalize the result of their cross product in Equation~\ref{eqn:v} to get $\hat{v}$.

By combining Equations~\ref{eqn:v} and~\ref{eqn:u} with Equation~\ref{eqn:n}, we can fully compute a new $\hat{n}$ that can be used for lighting.

\subsection{Automatic Differentiation}
\label{sec:ad}

The derivatives in Section~\ref{sec:updatednormals} are being used to compute the Jacobian of the offset function $\in~\mathbb{R}^{3\times 3}$ based on an input position $\in~\mathbb{R}^3$. Given that the outputs are greater in number than the inputs, we opt for forward-mode automatic differentiation.

The derivatives will be used in a shader, which has limited language capabilities and resources. Rather than creating structs in GLSL to create dual numbers and defining GLSL functions to operate on them, we instead differentiate at compile time. We express the offset function in a host language outside of the shader, and from the host language, generate GLSL source code for the shader that computes both the offset and its derivatives. For each operation in the computation graph, we can generate one line of GLSL to compute and store the output of the computation, plus an additional line storing the derivative of that graph node with respect to each independent variable. In practice, we output more condensed code, as we are able to compact multiple computation graph nodes up to a specified depth before outputting GLSL, as we find this easier to inspect and debug. Listing~\ref{lst:ad} shows an example of this, with and without condensed output.

\lstdefinelanguage{GLSL}%
{%
	morekeywords={%
		false,FALSE,NULL,true,TRUE,%
		__LINE__,__FILE__,__VERSION__,GL_core_profile,GL_es_profile,GL_compatibility_profile,%
		precision,highp,mediump,lowp,%
		break,case,continue,default,discard,do,else,for,if,return,switch,while,%
		void,bool,int,uint,float,double,vec2,vec3,vec4,dvec2,dvec3,dvec4,bvec2,bvec3,bvec4,ivec2,ivec3,ivec4,uvec2,uvec3,uvec4,mat2,mat3,mat4,mat2x2,mat2x3,mat2x4,mat3x2,mat3x3,mat3x4,mat4x2,mat4x3,mat4x4,dmat2,dmat3,dmat4,dmat2x2,dmat2x3,dmat2x4,dmat3x2,dmat3x3,dmat3x4,dmat4x2,dmat4x3,dmat4x4,sampler1D,sampler2D,sampler3D,image1D,image2D,image3D,samplerCube,imageCube,sampler2DRect,image2DRect,sampler1DArray,sampler2DArray,image1DArray,image2DArray,samplerBuffer,imageBuffer,sampler2DMS,image2DMS,sampler2DMSArray,image2DMSArray,samplerCubeArray,imageCubeArray,sampler1DShadow,sampler2DShadow,sampler2DRectShadow,sampler1DArrayShadow,sampler2DArrayShadow,samplerCubeShadow,samplerCubeArrayShadow,isampler1D,isampler2D,isampler3D,iimage1D,iimage2D,iimage3D,isamplerCube,iimageCube,isampler2DRect,iimage2DRect,isampler1DArray,isampler2DArray,iimage1DArray,iimage2DArray,isamplerBuffer,iimageBuffer,isampler2DMS,iimage2DMS,isampler2DMSArray,iimage2DMSArray,isamplerCubeArray,iimageCubeArray,atomic_uint,usampler1D,usampler2D,usampler3D,uimage1D,uimage2D,uimage3D,usamplerCube,uimageCube,usampler2DRect,uimage2DRect,usampler1DArray,usampler2DArray,uimage1DArray,uimage2DArray,usamplerBuffer,uimageBuffer,usampler2DMS,uimage2DMS,usampler2DMSArray,uimage2DMSArray,usamplerCubeArray,uimageCubeArray,struct,%
		gl_BackColor,gl_BackLightModelProduct,gl_BackLightProduct,gl_BackMaterial,gl_BackSecondaryColor,gl_ClipDistance,gl_ClipPlane,gl_ClipVertex,gl_Color,gl_DepthRange,gl_DepthRangeParameters,gl_EyePlaneQ,gl_EyePlaneR,gl_EyePlaneS,gl_EyePlaneT,gl_Fog,gl_FogCoord,gl_FogFragCoord,gl_FogParameters,gl_FragColor,gl_FragCoord,gl_FragData,gl_FragDepth,gl_FrontColor,gl_FrontFacing,gl_FrontLightModelProduct,gl_FrontLightProduct,gl_FrontMaterial,gl_FrontSecondaryColor,gl_InstanceID,gl_Layer,gl_LightModel,gl_LightModelParameters,gl_LightModelProducts,gl_LightProducts,gl_LightSource,gl_LightSourceParameters,gl_MaterialParameters,gl_ModelViewMatrix,gl_ModelViewMatrixInverse,gl_ModelViewMatrixInverseTranspose,gl_ModelViewMatrixTranspose,gl_ModelViewProjectionMatrix,gl_ModelViewProjectionMatrixInverse,gl_ModelViewProjectionMatrixInverseTranspose,gl_ModelViewProjectionMatrixTranspose,gl_MultiTexCoord0,gl_MultiTexCoord1,gl_MultiTexCoord2,gl_MultiTexCoord3,gl_MultiTexCoord4,gl_MultiTexCoord5,gl_MultiTexCoord6,gl_MultiTexCoord7,gl_Normal,gl_NormalMatrix,gl_NormalScale,gl_ObjectPlaneQ,gl_ObjectPlaneR,gl_ObjectPlaneS,gl_ObjectPlaneT,gl_Point,gl_PointCoord,gl_PointParameters,gl_PointSize,gl_Position,gl_PrimitiveIDIn,gl_ProjectionMatrix,gl_ProjectionMatrixInverse,gl_ProjectionMatrixInverseTranspose,gl_ProjectionMatrixTranspose,gl_SecondaryColor,gl_TexCoord,gl_TextureEnvColor,gl_TextureMatrix,gl_TextureMatrixInverse,gl_TextureMatrixInverseTranspose,gl_TextureMatrixTranspose,gl_Vertex,gl_VertexID,%
		gl_MaxClipPlanes,gl_MaxCombinedTextureImageUnits,gl_MaxDrawBuffers,gl_MaxFragmentUniformComponents,gl_MaxLights,gl_MaxTextureCoords,gl_MaxTextureImageUnits,gl_MaxTextureUnits,gl_MaxVaryingFloats,gl_MaxVertexAttribs,gl_MaxVertexTextureImageUnits,gl_MaxVertexUniformComponents,%
		abs,acos,all,any,asin,atan,ceil,clamp,cos,cross,degrees,dFdx,dFdy,distance,dot,equal,exp,exp2,faceforward,floor,fract,ftransform,fwidth,greaterThan,greaterThanEqual,inversesqrt,length,lessThan,lessThanEqual,log,log2,matrixCompMult,max,min,mix,mod,noise1,noise2,noise3,noise4,normalize,not,notEqual,outerProduct,pow,radians,reflect,refract,shadow1D,shadow1DLod,shadow1DProj,shadow1DProjLod,shadow2D,shadow2DLod,shadow2DProj,shadow2DProjLod,sign,sin,smoothstep,sqrt,step,tan,texture1D,texture1DLod,texture1DProj,texture1DProjLod,texture2D,texture2DLod,texture2DProj,texture2DProjLod,texture3D,texture3DLod,texture3DProj,texture3DProjLod,textureCube,textureCubeLod,transpose,%
		rgb
	},
	sensitive=true,%
	morecomment=[s]{/*}{*/},%
	morecomment=[l]//,%
	morestring=[b]",%
	morestring=[b]',%
	moredelim=*[directive]\#,%
	moredirectives={define,defined,elif,else,if,ifdef,endif,line,error,ifndef,include,pragma,undef,warning,extension,version}%
}[keywords,comments,strings,directives]%

\begin{lstlisting}[language=GLSL, caption={Generated GLSL code for the Jacobian of the offset function \unexpanded{$f([x~~y~~z]^T)~=~[0.5\sin(0.005t + 2y)~~0~~0]^T$}, with and without condensing the number of intermediate variables in the output. The automatic output has been manually formatted and intermediate variables have been manually renamed for readability.}, label={lst:ad}, float, texcl=false, mathescape=false,]
// One node per variable:
float v1 = millis * 0.005;
float v2 = position.y * 2.0;
float d_v2_by_d_y = 2.0 * 1.0;
float v3 = v1 + v2;
float d_v3_by_d_y = 0.0 + d_v2_by_d_y;
float v4 = sin(v3);
float d_v4_by_d_y =
  cos(v3) * d_v3_by_d_y;
float v5 = v4 * 0.5;
float d_v5_by_d_y = 0.5 * d_v4_by_d_y;
vec3 offset = vec3(v5, 0.0, 0.0);
vec3 d_offset_by_d_y =
  vec3(d_v5_by_d_y, 0.0, 0.0); 
mat3 jacobian = mat3(
  vec3(0.0),
  d_offset_by_d_y,
  vec3(0.0)
);

// Condensed:
vec3 offset = vec3(
  sin(time * 0.005 + position.y * 2.0)
    * 0.5,
  0.0,
  0.0
);
vec3 d_offset_by_d_y = vec3(
  0.5 * cos(
    time * 0.005 + position.y * 2.0
  )) * (0.0 + (2.0 * 1.0)),
  0.0,
  0.0
);
mat3 jacobian = mat3(
  vec3(0.0),
  d_offset_by_d_y,
  vec3(0.0)
);
\end{lstlisting}

\section{Implementation}

Targeting the creative coding community, our implementation comes in the form of the library \textit{p5.warp}: a plugin for the web graphics library p5.js~\cite{p5.js} written in JavaScript and GLSL. Artists specify an offset function $f$ in Javascript through our API, a shader generator using the builder design pattern. The library provides builder methods corresponding to mathematical operators and functions, allowing a computation graph to be built with concise syntax. The builder implements static forward-mode automatic differentiation by outputting GLSL code containing expressions that compute both $f$ and its Jacobian $J$.

The artist does not need to write any GLSL and only needs to interact with our builder API. Artists are free to use loops, make temporary variables, or create and call functions while interacting with the builder. These JavaScript constructs act as macros with respect to the generated shader code, evaluated at the shader's compile time to generate static GLSL. The API also provides access to inputs such as the current time in milliseconds, the mouse position, or the canvas size, which get turned into shader uniforms. Artists may define a warp in either model or world space. Listing~\ref{lst:twist} shows an example of a twist warp that animates back and forth over time, using the \lstinline{millis} uniform to access time and using a JavaScript function as a macro.

The output of the builder is spliced into a vertex shader, where the normal-updating math from Section~\ref{sec:formulation} is implemented. The main subroutine of the shader for a warp function defined in model space is shown in Listing~\ref{lst:shader}.

\definecolor{param}{rgb}{.6,0,.6}
\lstdefinelanguage{JavaScript}{
  keywords={const, break, case, catch, continue, debugger, default, delete, do, else, finally, for, function, if, in, instanceof, new, return, switch, this, throw, try, typeof, var, void, while, with},
  ndkeywords={glsl, millis, position},
  ndkeywordstyle=\color{param},
  morecomment=[l]{//},
  morecomment=[s]{/*}{*/},
  morestring=[b]',
  morestring=[b]",
  sensitive=true
}
\begin{lstlisting}[language=JavaScript, caption={An example of defining a twist about the horizontal axis in p5.warp.}, label={lst:twist}, float]
const twist = createWarp(
  function({
    glsl,
    millis,
    position
  }) {
    function rotateX(pos, angle) {
      const sa = glsl.sin(angle);
      const ca = glsl.cos(angle);
      return glsl.vec3(
        pos.x(),
        pos.y()
          .mult(ca)
          .sub(pos.z().mult(sa)),
        pos.y()
          .mult(sa)
          .add(pos.z().mult(ca))
      );
    };
    
    const rotated = rotateX(
      position,
      position.x()
        .mult(0.02)
        .mult(millis.div(1000).sin())
    );
    return rotated.sub(position);
  }
);
\end{lstlisting}

\begin{lstlisting}[language=GLSL, caption={A vertex shader snippet calculating an updated normal.}, label={lst:shader}, float, texcl=false, mathescape=false,]
void main() {
  // Start from attributes
  vec3 position = aPosition;
  vec3 normal = aNormal;

  // Splice in auto-generated code.
  // This defines `vec3 offset`, and
  // three `vec3`s for each column of
  // the Jacobian: `dodx`, `dody`,
  // and `dodz`
  ${outputOffsetAndDerivatives()}
  
  position += offset;
  vec3 w =
    (normal.y == 0. && normal.z == 0.)
      ? vec3(0., 1., 0.)
      : vec3(1., 0., 0.);
  vec3 v =
    normalize(cross(w, normal));
  vec3 u = cross(v, normal);
  mat3 jacobian =
    mat3(dodx, dody, dodz);
  normal = normalize(cross(
    u + jacobian * u,
    v + jacobian * v
  ));

  // Apply camera transforms
  // and output
  gl_Position =
    uP * uVM * vec4(position, 1.);
  
  // Pass on to fragment shader
  vNormal = uN * normal;
}
\end{lstlisting}

\section{Results}

\begin{figure}[h]
  \centering

    \begin{subfigure}[b]{0.22\textwidth}
        \includegraphics[trim={5cm 10cm 5cm 10cm},clip,width=\textwidth]{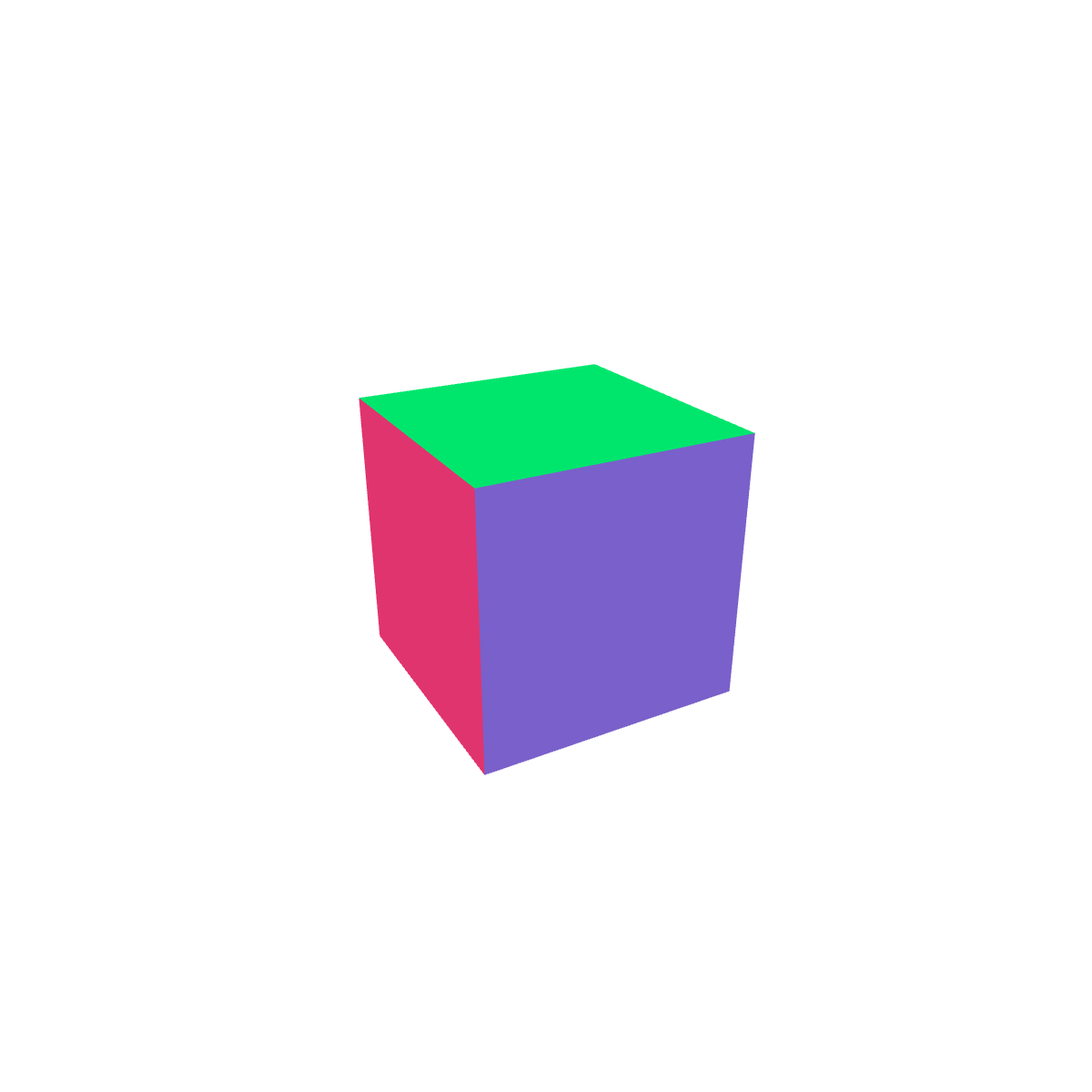}
        \caption{Input}
    \end{subfigure}
    \begin{subfigure}[b]{0.22\textwidth}
        \includegraphics[trim={5cm 10cm 5cm 10cm},clip,width=\textwidth]{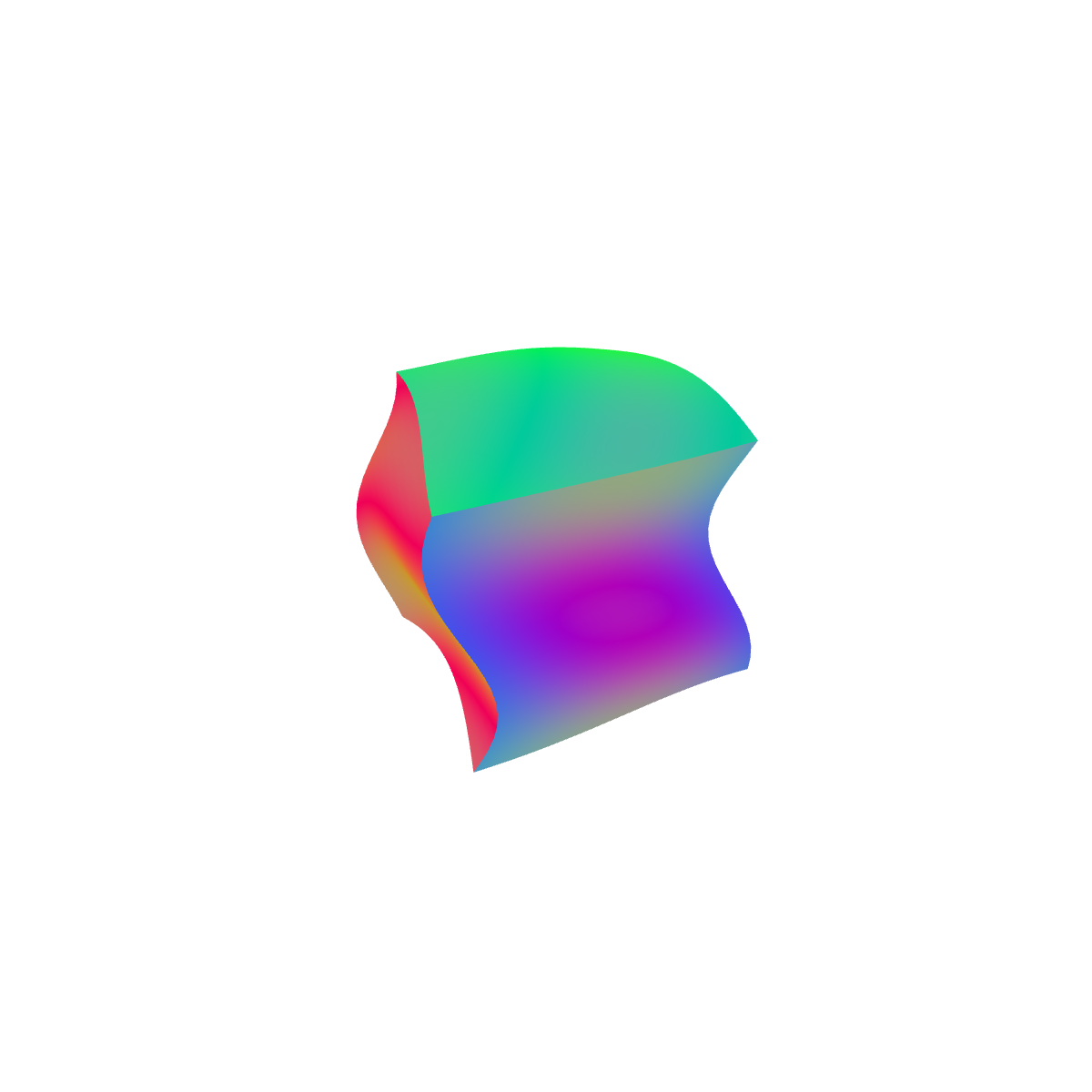}
        \caption{Warped}
    \end{subfigure}
  
   \caption{\label{fig:cube}
     A cube~(a) passed through a warp made with sine waves in each axis~(b).}
\end{figure}

\begin{figure}
  \centering

    \begin{subfigure}[t]{0.15\textwidth}
        \includegraphics[trim={10cm 12cm 10cm 12cm},clip,width=\textwidth]{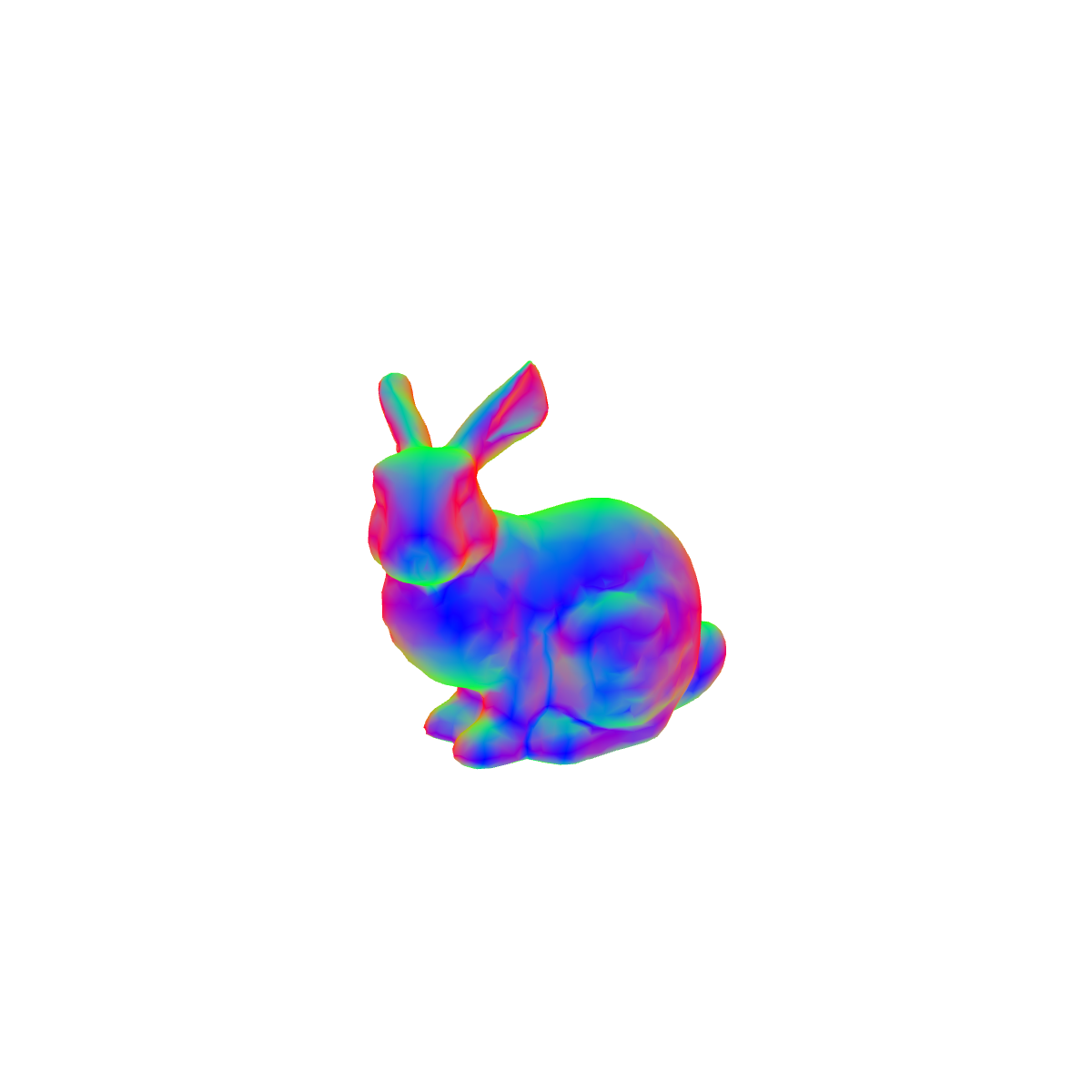}
        \caption{Input~\cite{bunny}}
    \end{subfigure}
    \begin{subfigure}[t]{0.15\textwidth}
        \includegraphics[trim={10cm 12cm 10cm 12cm},clip,width=\textwidth]{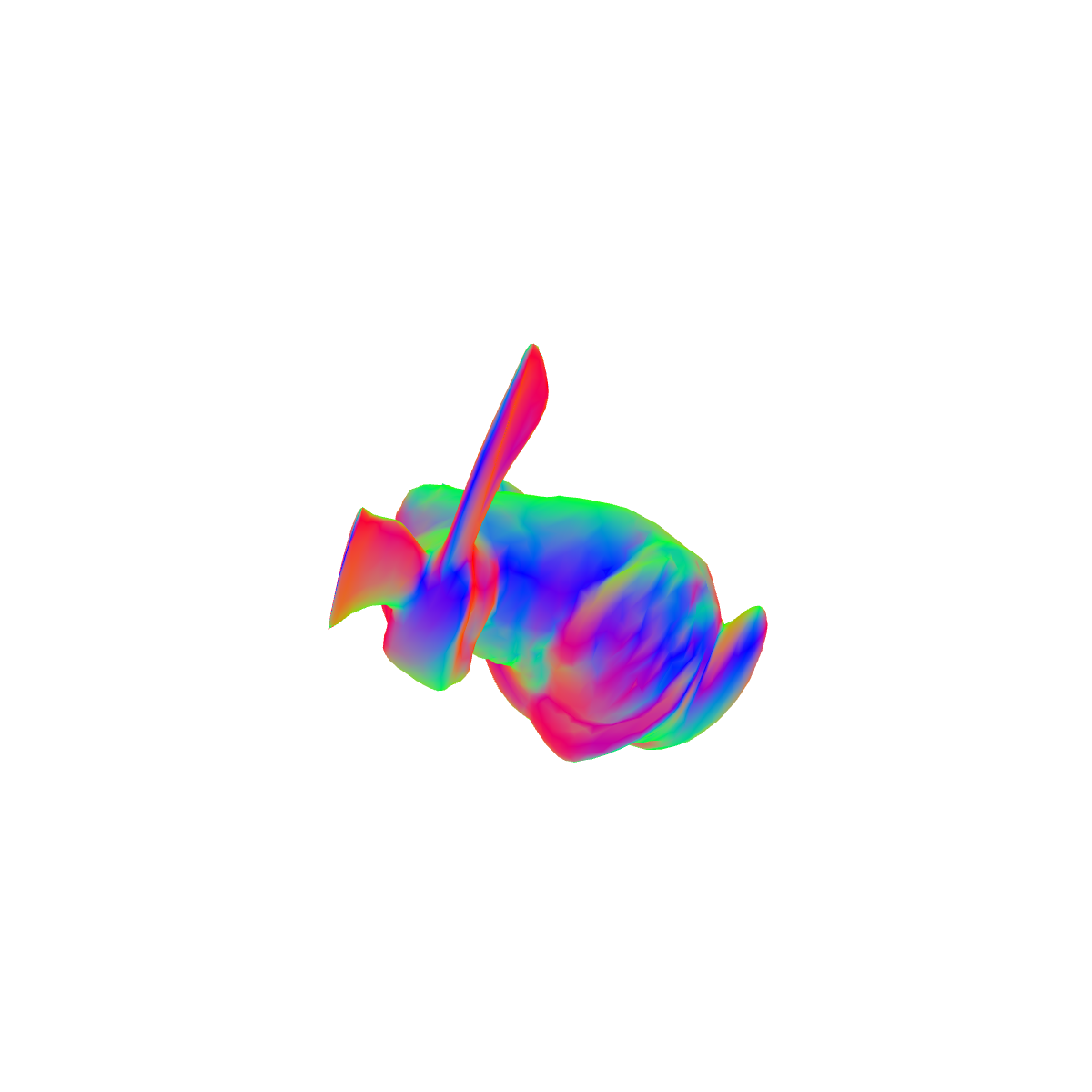}
        \caption{Warped with a twist}
    \end{subfigure}
    \begin{subfigure}[t]{0.15\textwidth}
        \includegraphics[trim={10cm 12cm 10cm 12cm},clip,width=\textwidth]{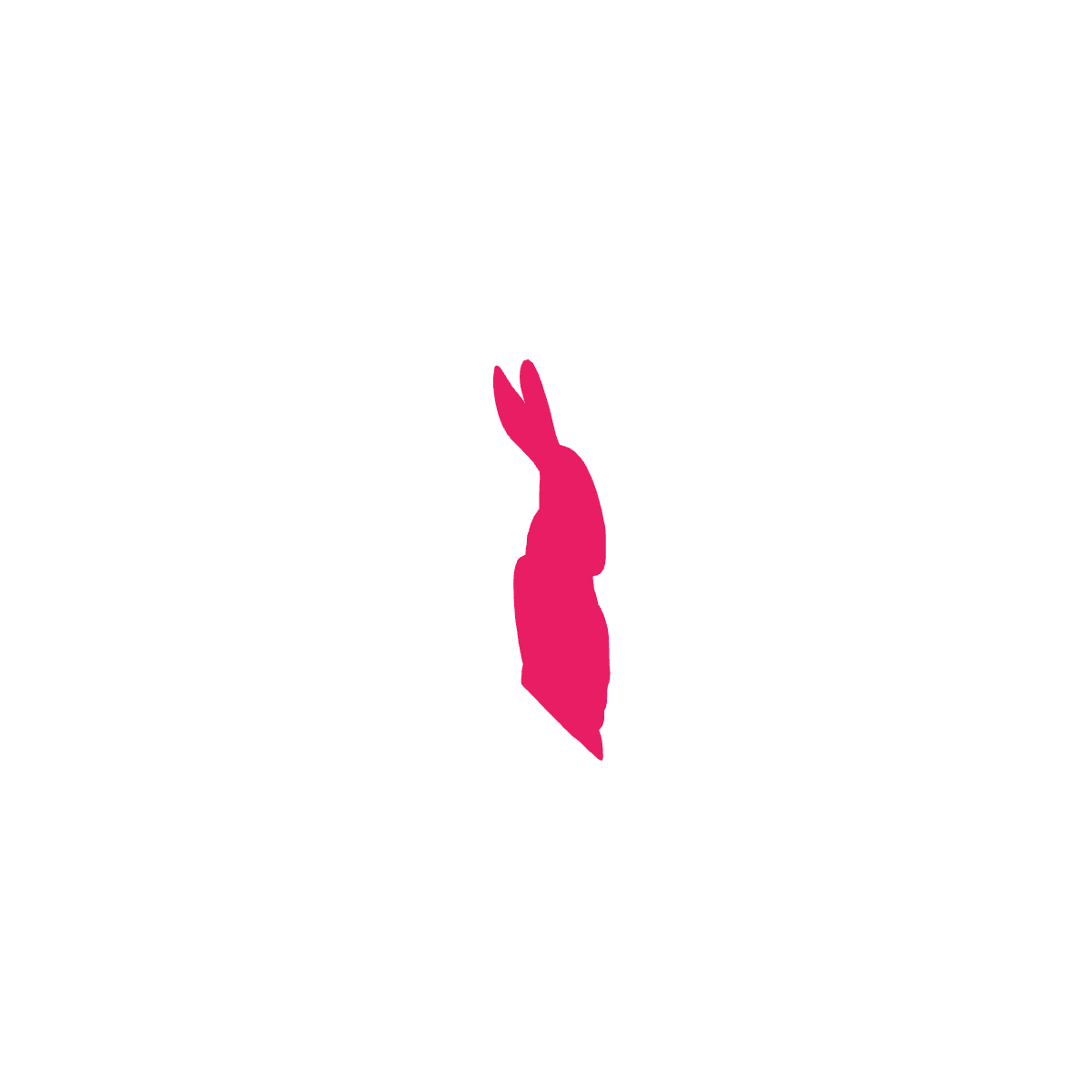}
        \caption{\label{fig:bunny-flat}Warped by flattening}
    \end{subfigure}
  
   \caption{\label{fig:bunny}
     A bunny~(a) retaining correct normals when passed through warps that twist about the X axis~(b) and even after being flattened into the YZ plane~(c).}
\end{figure}

\begin{figure}
    \centering
  \includegraphics[trim={3cm 0cm 0cm 4cm},clip,width=1.2in]{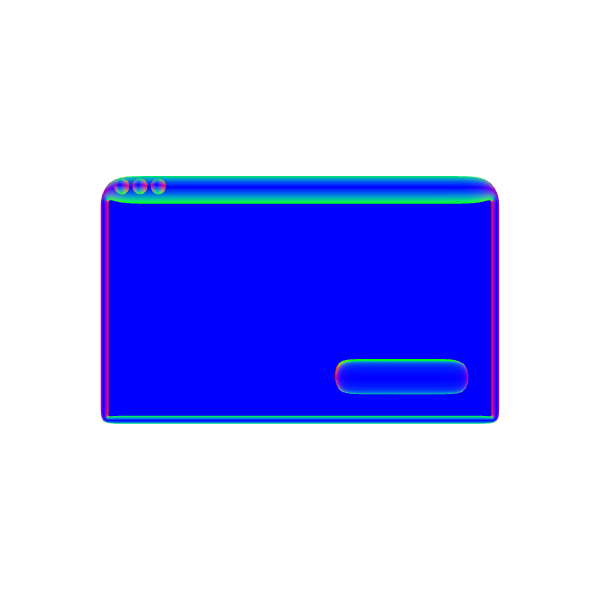}
  \includegraphics[trim={3cm 0cm 0cm 4cm},clip,width=1.2in]{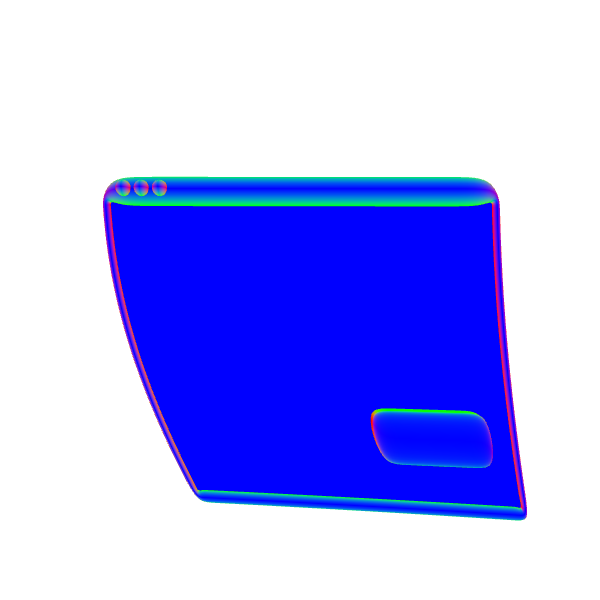}
  \includegraphics[trim={3cm 0cm 0cm 4cm},clip,width=1.2in]{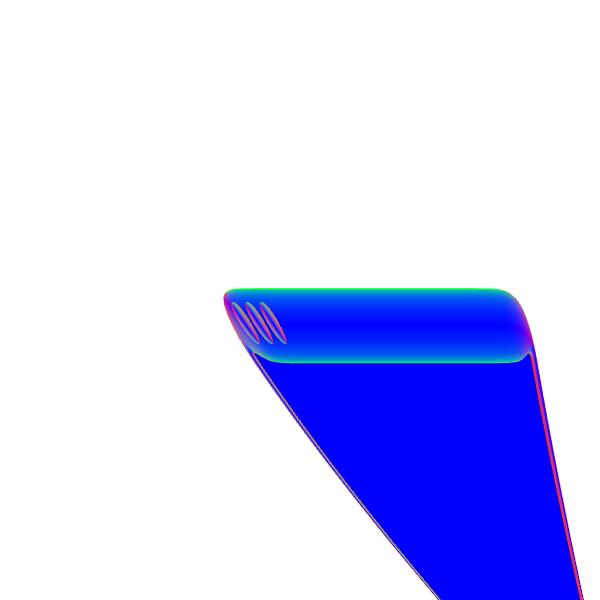}
  \includegraphics[trim={3cm 0cm 0cm 4cm},clip,width=1.2in]{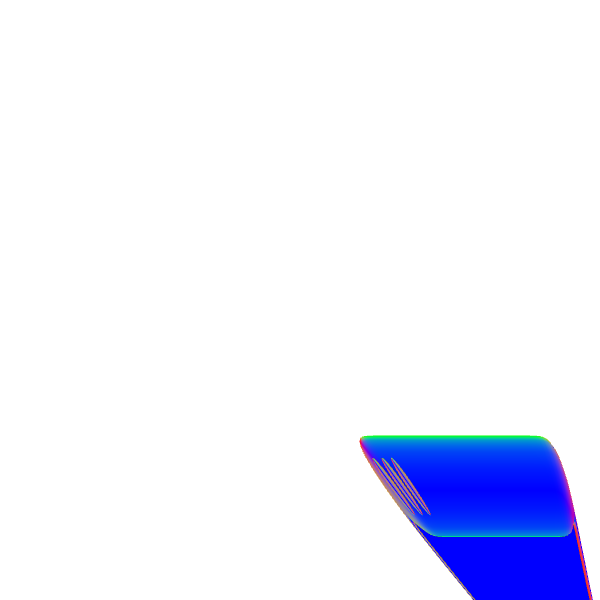}
  \caption{Frames of an animated warp inspired by the ``genie'' animation while minimizing a window introduced in Mac OS X.}
  \label{fig:genie}
\end{figure}

\begin{figure*}
  \centering

    \begin{subfigure}[t]{0.59\textwidth}
        \includegraphics[trim={0cm 5cm 0cm 0cm},clip,width=\textwidth]{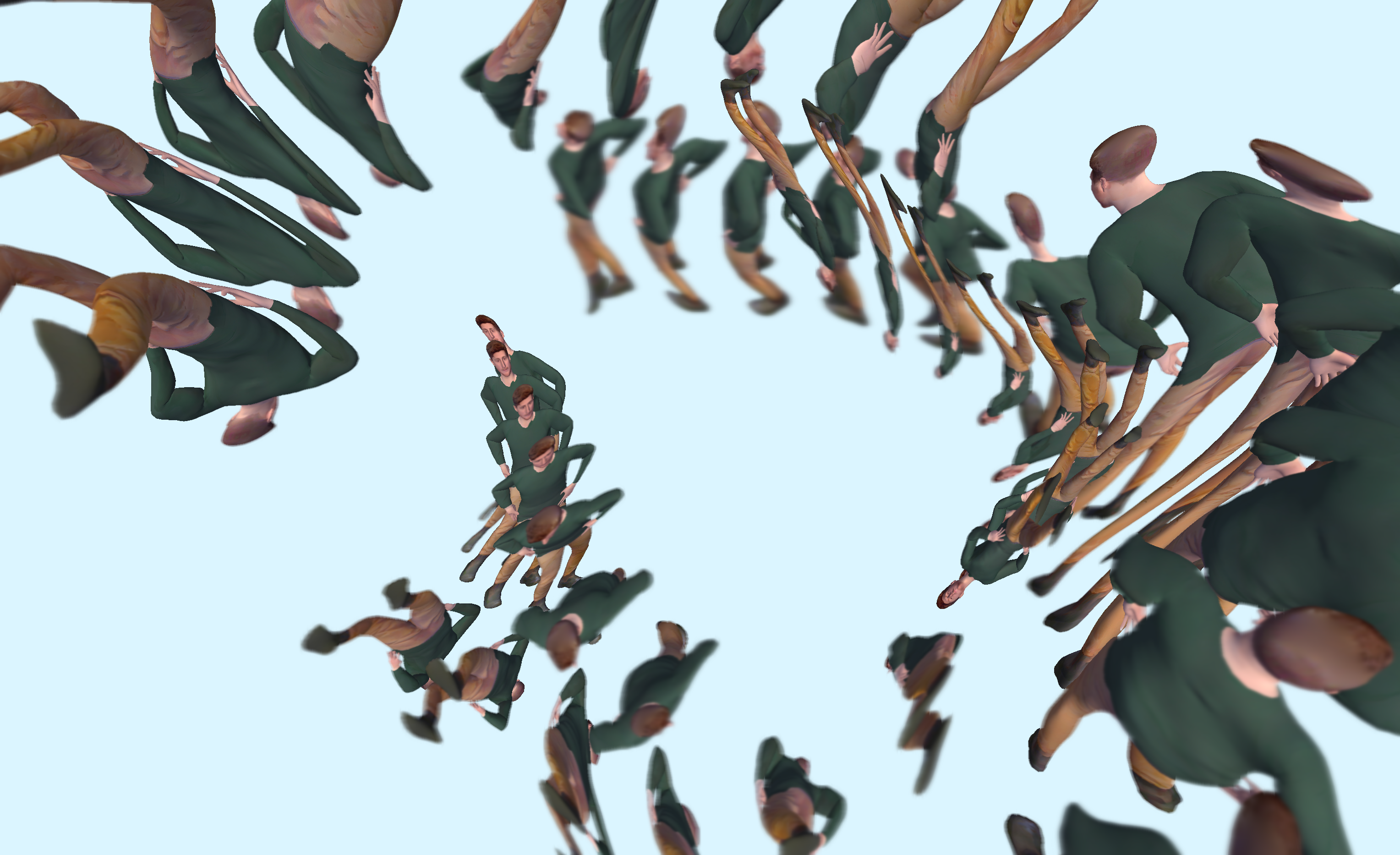}
        \caption{Repeated, warped copies of a 3D scan of a person, creating a surreal composition.}
    \end{subfigure}
    \hspace{1em}
    \begin{subfigure}[t]{0.33\textwidth}
        \includegraphics[width=\textwidth]{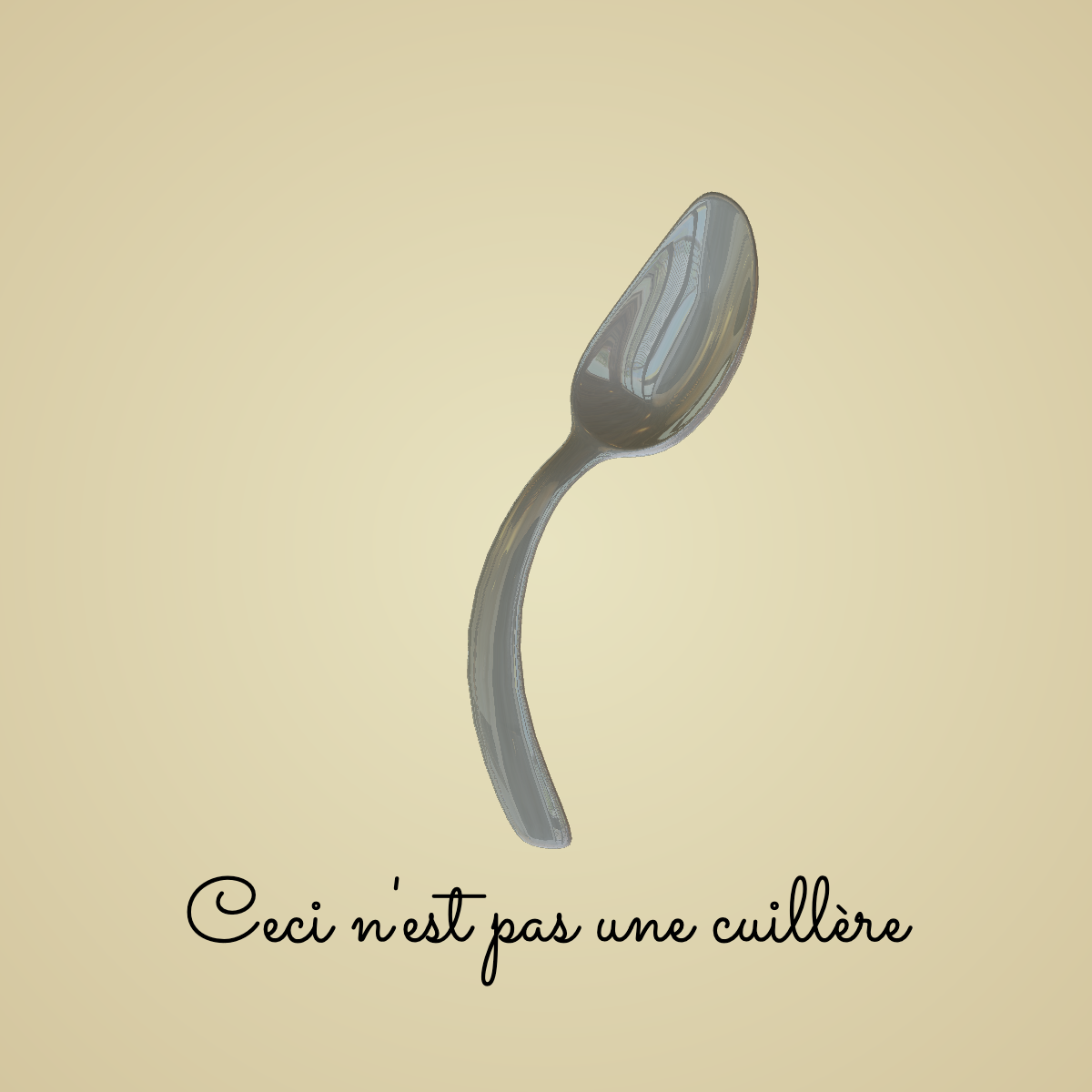}
        \caption{A warped spoon, playing on \em The Matrix \em and Ren\'e Magritte's \em The Treachery of Images. \em}
    \end{subfigure}
    
    \begin{subfigure}[t]{0.33\textwidth}
        \includegraphics[width=\textwidth]{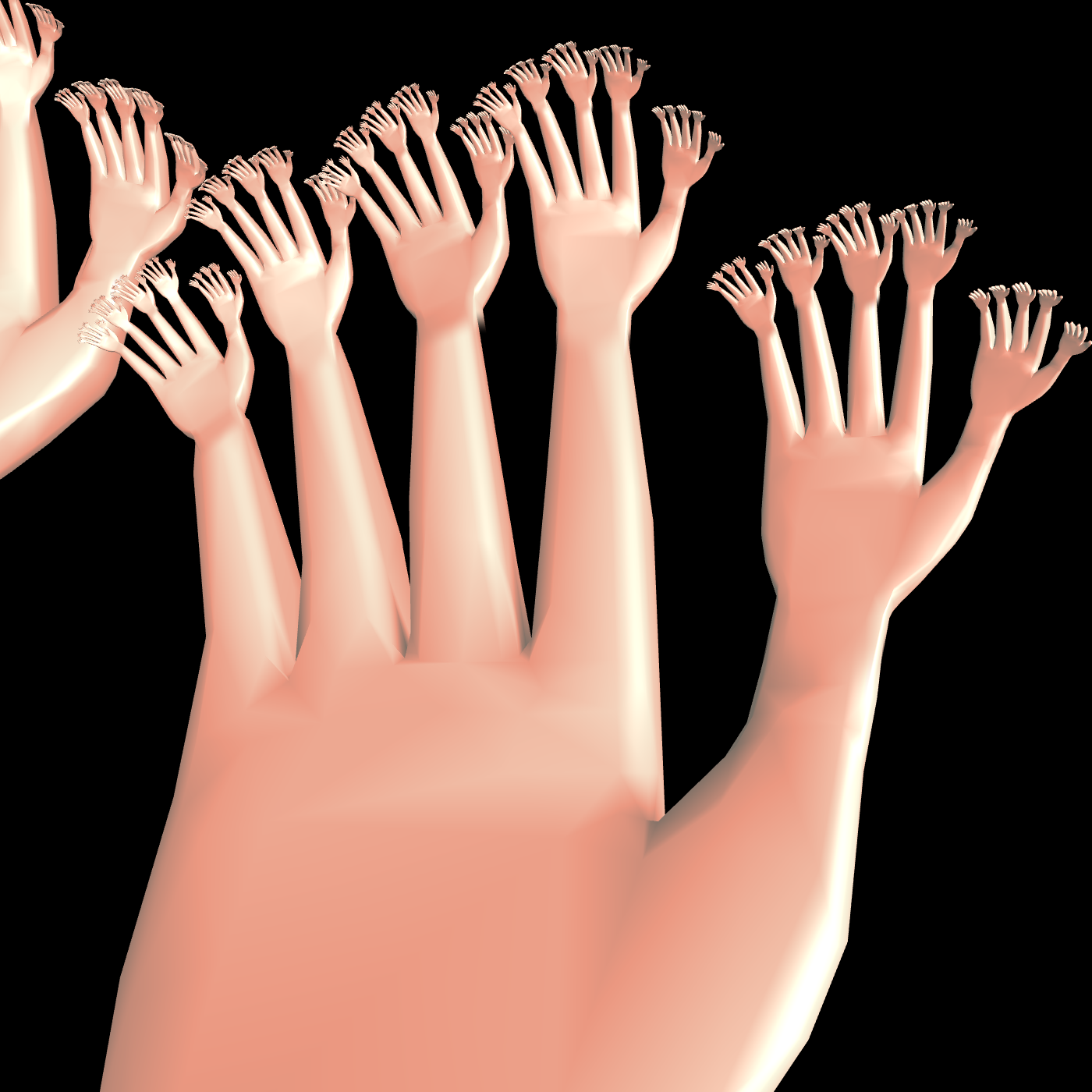}
        \caption{A recursive hand structure using world-space warping to make fingers wiggle.}
    \end{subfigure}
    \hspace{1em}
    \begin{subfigure}[t]{0.59\textwidth}
        \includegraphics[trim={0cm -5cm 0cm 0cm},width=\textwidth]{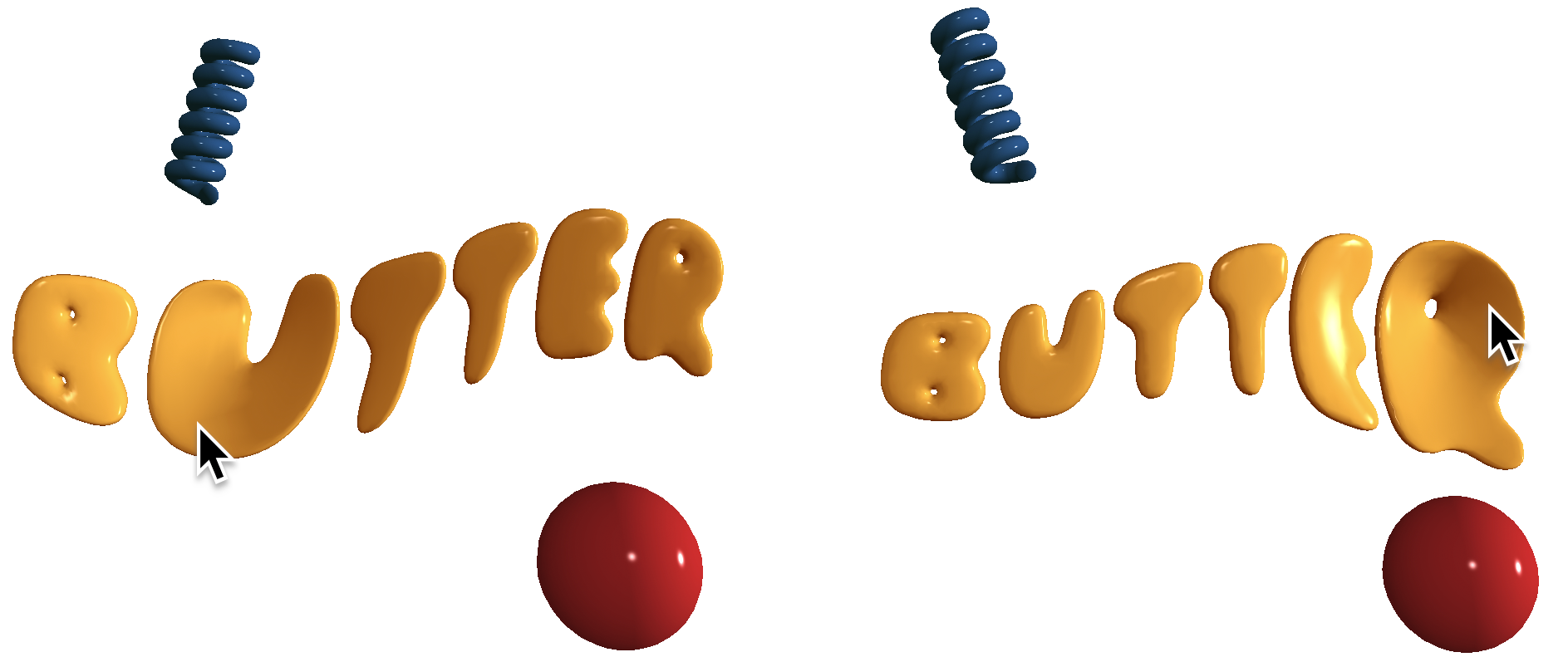}
        \caption{A bulge effect added to 3D text based on the mouse location.}
    \end{subfigure}

    \vspace{1em}
    \begin{subfigure}[t]{0.95\textwidth}
        \includegraphics[trim={0cm 18cm 0cm 18cm},clip,width=0.32\textwidth]{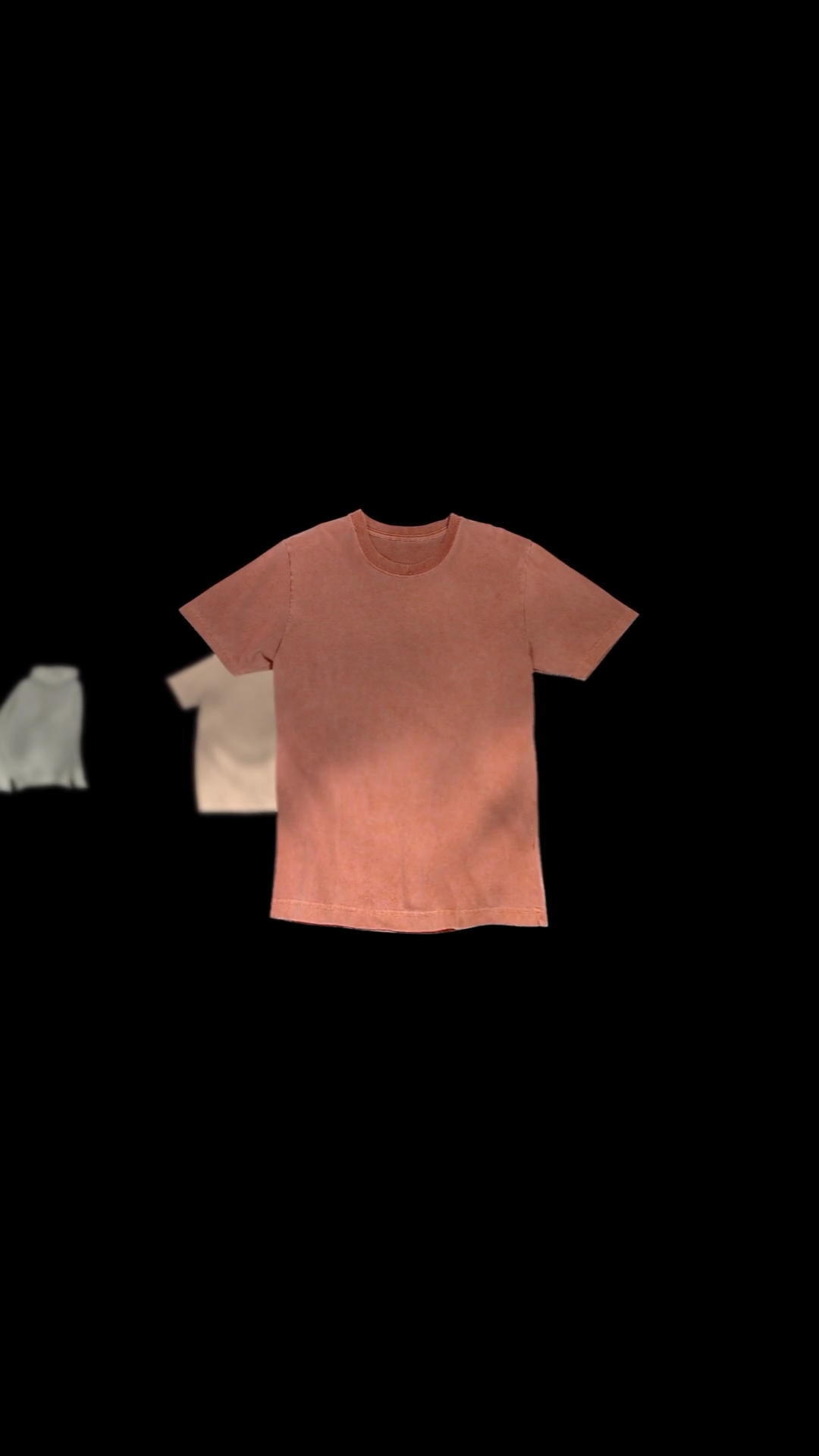}
        \includegraphics[trim={0cm 18cm 0cm 18cm},clip,width=0.32\textwidth]{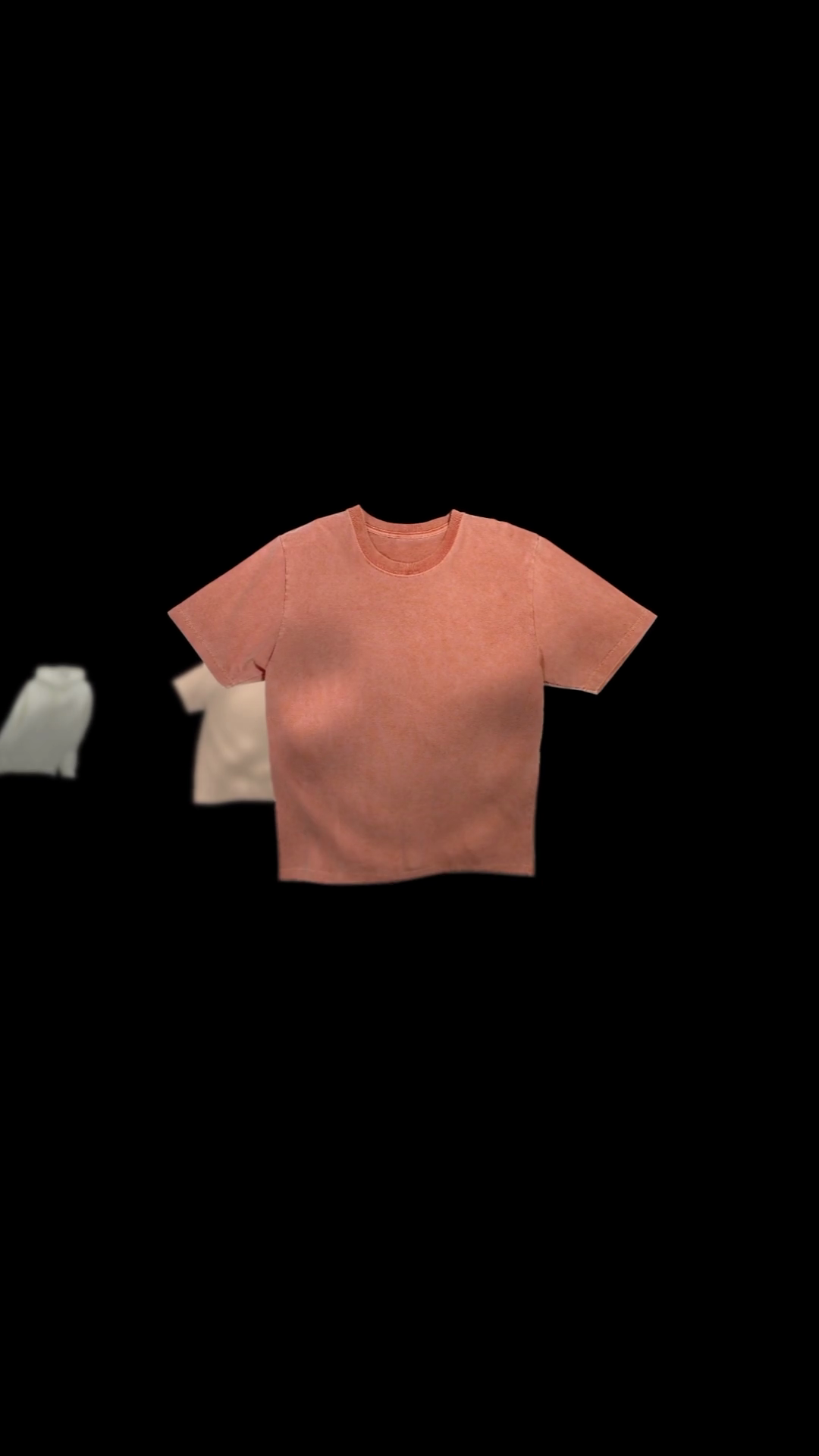}
        \includegraphics[trim={0cm 18cm 0cm 18cm},clip,width=0.32\textwidth]{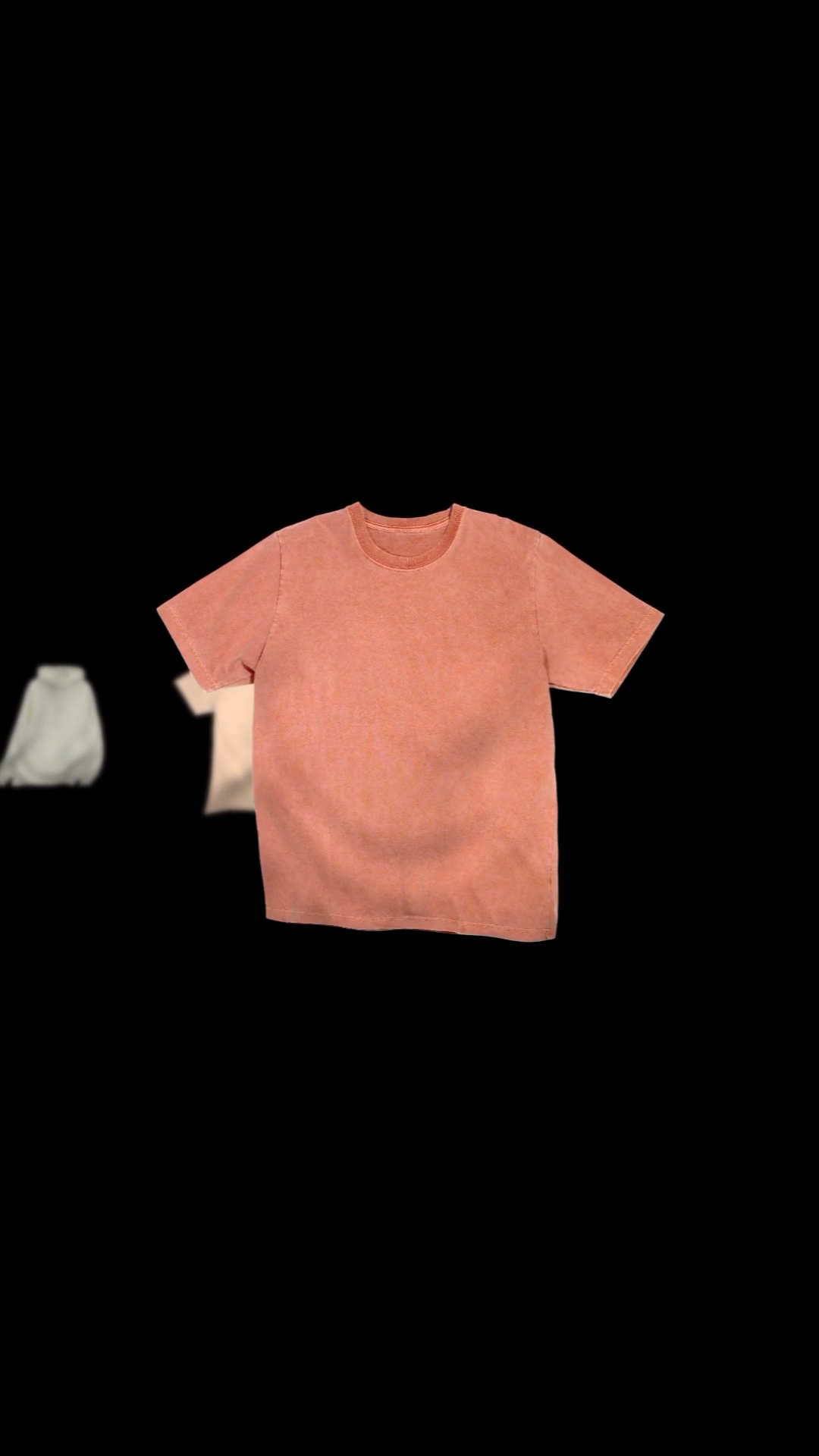}
        \caption{Pretend cloth physics applied via a warp to a rotating carousel of shirts, where each is a textured plane.}
    \end{subfigure}
  
   \caption{\label{fig:examples}
        Examples of domain warping using p5.warp in different sketches, using lighting calculations in their fragment shaders that make use of updated normals.}
\end{figure*}

\paragraph{Visuals.} Figure~\ref{fig:teaser} shows frames of an animated warp applied to an airplane model, giving it a twisting motion with cartoon squash and stretch. Examples of other warps applied to models can be seen in figures~\ref{fig:spherewarp},~\ref{fig:cube},~\ref{fig:bunny}~and~\ref{fig:genie}. For easier inspection of accuracy, they have been visualized with a fragment shader that outputs the absolute value of the updated normal as the color. Figure~\ref{fig:examples} shows warps applied to models as part of larger compositions, with lighting calculations done in their fragment shaders.

%
%

\paragraph{Performance.} As a test of a typical use case, we measured the performance of the animation shown in Figure~\ref{fig:teaser}. It runs p5.warp on a mesh with around 14,000 vertices using a warp with 150 nodes in its computation graph, rendering onto a 1200$\times$1200 pixel canvas. It runs at 60 frames per second, the imposed browser limit on a 60Hz display, on a 2021 MacBook Pro with an M1 Pro chip, a 2015 Intel MacBook Pro with integrated GPU, and a 2021 Motorola Edge Android phone in Google Chrome.

We stress-tested our method by increasing both the number of vertices in the model and the number of nodes in the computation graph. The results are listed in Table~\ref{tbl:perf}. When increasing the vertex count of the model, the MacBooks maintained 60fps, and the tab on the phone crashed before losing 60fps, suggesting that hardware memory limitations may be the primary bottleneck for model size rather than the performance of p5.warp with standard warp complexity. We also experimented with increasing the warp complexity by stacking random sine waves until reaching a target number of nodes in the computation graph. This had a more noticeable effect on frame rates: the phone begins to struggle at 1,500 nodes, and the 2015 MacBook begins to show signs of slowing down at 3,000. We consider these to be acceptable limits, as our manually coded warps have a smaller computation graph by a factor of 10, leaving sufficient performance buffer room.

\begin{table*}[htp]
\small
\begin{center}
\begin{tabularx}{\textwidth}{l|X|X|X}
& 2021 M1 Pro MacBook Pro &  2015 Integrated GPU Intel MacBook Pro & 2021 Motorola Edge\\
\hline
14,000 vertices, 150 nodes & 60fps & 60fps & 60fps \\
500,000 vertices, 150 nodes & 60fps & 60fps & Crash \\
14,000 vertices, 1,500 nodes & 60fps & 60fps & 26fps \\
14,000 vertices, 3,000 nodes & 60fps & 50fps & 10fps \\
\end{tabularx}
\end{center}
\caption{Frame rates achieved using our method on a variety of hardware, model sizes (number of vertices), and warp complexities (number of nodes in the computation graph.)}
\label{tbl:perf}
\end{table*}

\subsection{Discussion}

\begin{figure}[h]
  \centering

    \begin{subfigure}[b]{0.15\textwidth}
        \includegraphics[trim={10cm 10cm 10cm 10cm},clip,width=\textwidth]{img/cube.png}
        \caption{Input}
    \end{subfigure}
    \begin{subfigure}[b]{0.15\textwidth}
        \includegraphics[trim={10cm 10cm 10cm 10cm},clip,width=\textwidth]{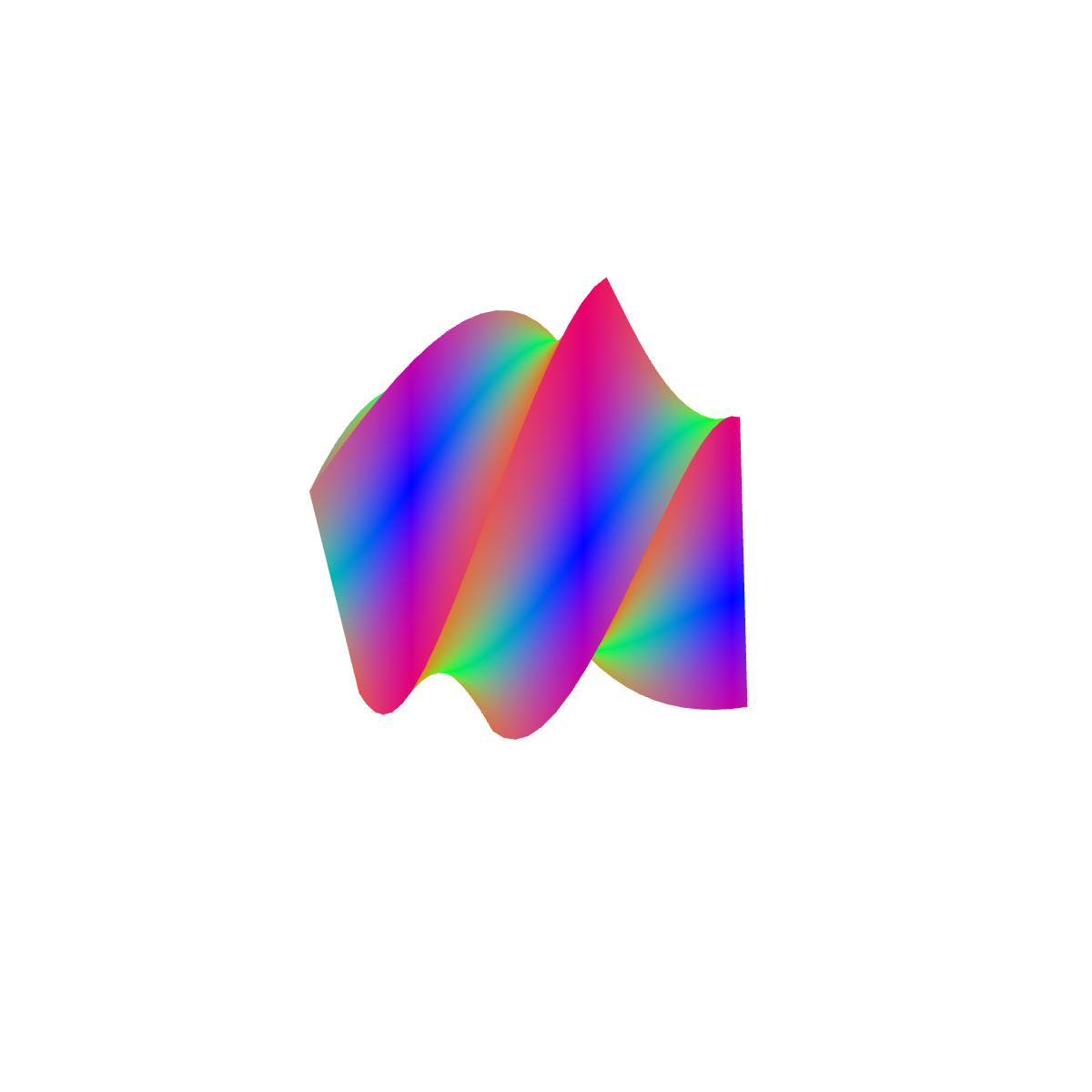}
        \caption{Warped, 20 edges/side}
    \end{subfigure}
    \begin{subfigure}[b]{0.15\textwidth}
        \includegraphics[trim={10cm 10cm 10cm 10cm},clip,width=\textwidth]{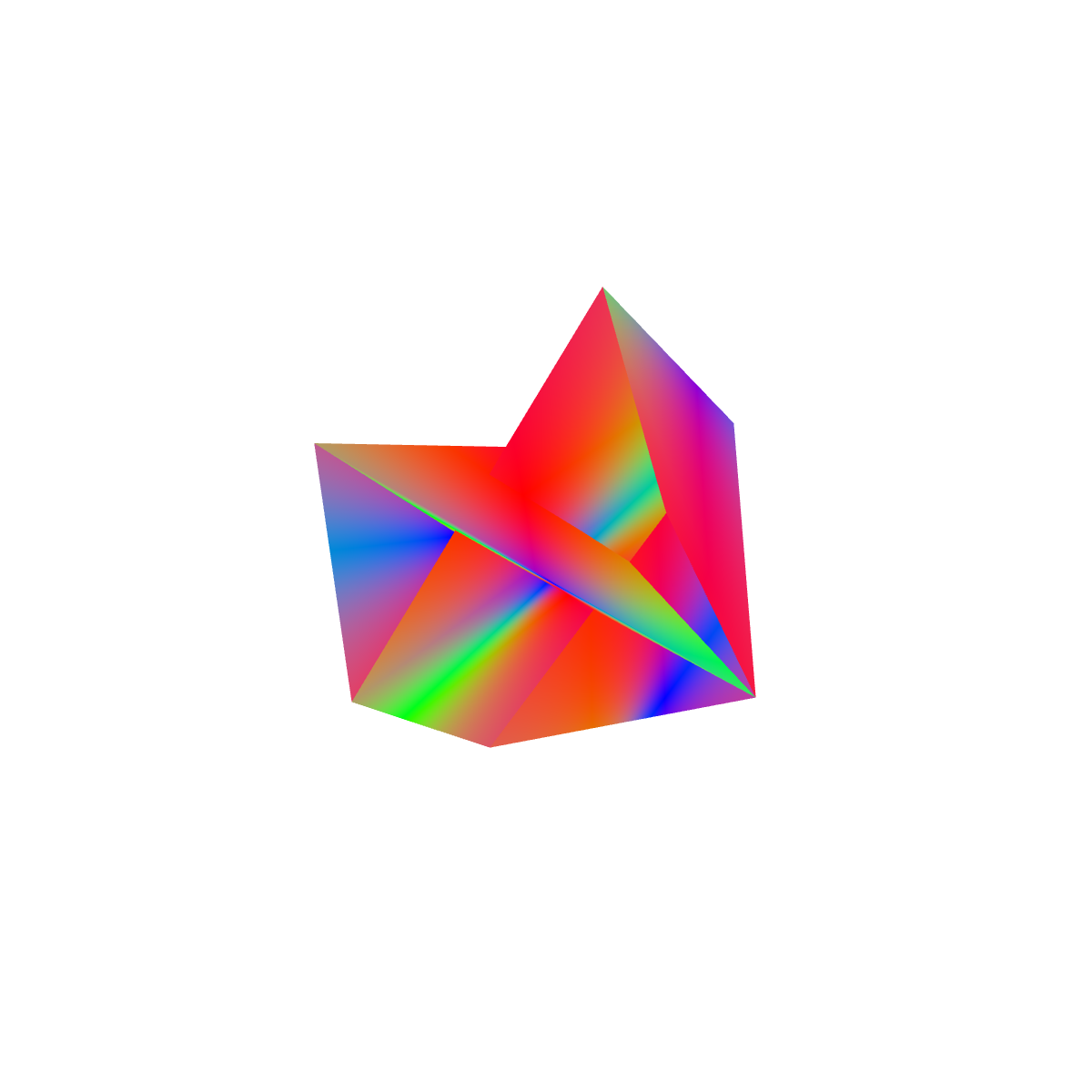}
        \caption{Warped, one edge/side}
    \end{subfigure}
  
   \caption{\label{fig:resolution}
     A cube~(a) passed through a twist warp when the cube has been subdivided into 20 edges per side~(b) and one edge per side~(c). Without sufficient polygon detail, the linear interpolation between vertices fails to capture the full warp, producing self-intersections and less accurate normals interpolated over faces.}
\end{figure}

\paragraph{Model resolution.} Given two points $\vec{p}$ and $\vec{q}$ connected by a straight edge on a mesh $M$, the ideal form of the domain warped mesh has $\vec{p'}=w(\vec{p})$ and $\vec{q'}=w(\vec{q})$ connected by a curve $C(t) = w(t\vec{p} + (1-t)\vec{q}),~t\in[0,1]$. Since the outputted mesh in WebGL will still connect $\vec{p'}$ and $\vec{q'}$ with the straight line segment $L(t) = tw(\vec{p}) + (1-t)w(\vec{q})$, the visual quality of a deformed mesh depends highly on the distance between the ideal curve $C(t)$ and its approximation $L(t)$. The effect of model resolution on a high frequency warp is shown in Figure~\ref{fig:resolution}.

\paragraph{Continuum approximation.} Warp functions are defined over $\mathbb{R}^3$ as a continuum without knowledge of the true shape of the model. While this method has been used to implement efficient hand-animated \textit{imitations} of physical phenomena such as soft body springiness or cloth in the wind, this method is not suited for more accurate physics \textit{simulations} where forces propagate through the true shape of the mesh.

\paragraph{Degenerate offsets.} Rendering is rarely affected by the cases where the equations in Section~\ref{sec:formulation} would produce zero-magnitude vectors as normals. Equation~\ref{eqn:n} breaks down if either operand of the cross product is a zero vector, or if both operands are equal to each other. The former case happens if, for some tangent vector $\hat{w}$, $J\hat{w}=-\hat{w}$, or in other terms, if $J$ has an eigenvalue of -1. Rearranging the equation for the latter case, we find it is a special case of the former case:
\begin{align*}
    \hat{u} + J \hat{u} &= \hat{v} + J \hat{v}\\
    J(\hat{u} - \hat{v}) &= -(\hat{u} - \hat{v})\\
    J\hat{w} &= -\hat{w},~~~~~~~~~~\hat{w} = \hat{u} - \hat{v}
\end{align*}

A warp where $\exists~\hat{w}~|~J\hat{w}=-\hat{w}$ is one that maps some axis to a constant. As an example, take $f(\vec{p})=-\vec{p} + \vec{k}$, which offsets any input to a constant in all axes, turning it into the point $\vec{k}$. Its Jacobian matrix is $-I_3$, which has an eigenvalue of -1.

The previous example choice of $f$ causes all normals to be degenarate, but by mapping all vertices to a single point, the mesh will have no surface area and no fragments will render, eliminating the problem. If all vertices are mapped to a line, which also has no surface area, similarly, no fragments will be rendered. If all vertices are mapped to a plane, as would happen with $f(\vec{p})=[-p_1, 0, 0]^T$, which flattens to the YZ plane, then some fragments will be rendered. Input normals tangent to the plane will have tangent vectors that get mapped to zero, being perpendicular to the plane. A face whose vertices are all such normals, if the normals are not askew from the face normal, is itself perpendicular to the YZ plane, so it will not get rendered, as it warps into a new face with zero area. Given a face where just some of its vertices have such normals, fragments in the middle of the face will have normals interpolated between some degenerate and some valid normals. In practice, this is not a problem: degenerate normals are zero vectors, and in the barycentric interpolation performed by WebGL, one or two components being zero simply reduces the magnitude of the weighted sum of the remaining vector components. Since the normal vector is normalized back to unit length in the fragment shader after interpolation, resulting normals will still be valid. Figure~\ref{fig:bunny-flat} shows the result of flattening to the YZ plane, where all resulting normals point in the X axis.

To summarize, offset functions that produce degenerate normals do not render fragments unless they collapse inputs to a plane. As long as faces whose vertices all have parallel normals form a surface that is itself perpenducular to those normals, plane-collapsing functions still produce correct results due to barycentric interpolation and normalization of normals.

\section{Conclusion and Future Work}

Domain warping, a common and versatile technique in creative coding, is difficult to use in 3D due to the need to recompute surface normals. We describe a method of computing updated normals that can be implemented in a vertex shader and present a library implementing such a shader, providing automatic generation of the derivatives is requires. One can then write any warp without worrying about incorrect lighting or parameter tuning, enabling artists to freely iterate.

With support for the WebGPU API now enabled by default in Google Chrome and the access to modern hardware capabilities it provides, there may be potential in the future to create new render pipelines using compute shaders that address current WebGL limitations. Our method currently lacks hardware-accelerated adaptive subdivision based on the warp function, which could perhaps be enabled with more GPU flexibility.

Finally, the ability to warp using arbitrary functions raises the human-computer interaction question of how to best to let artists sculpt abstract warp functions. Artists in the creative coding community are open to exploring math directly, but future work could include studies of more intuitive forms of warp function control.

\bibliographystyle{plainurl}
\bibliography{paper}

\section*{Index of Supplemental Materials}
Source code for p5.warp and its GLSL-building capabilities can respectively be found at:
\begin{itemize}
    \item \url{https://github.com/davepagurek/p5.warp}
    \item \url{https://github.com/davepagurek/glsl-autodiff}
\end{itemize}
A snapshot of both repositories is present in the supplemental material in a \lstinline{.zip} file.

\section*{Author Contact Information}

\hspace{-2mm}\begin{tabular}{p{0.5\textwidth}p{0.5\textwidth}}
Dave Pagurek van Mossel \newline
Butter Creatives \newline
\href{mailto:dave@davepagurek.com}{dave@davepagurek.com}
&

\end{tabular}

\end{document}